\documentclass[12pt]{revtex4}

\usepackage{graphicx}
\usepackage{dcolumn}
\usepackage{bm}
\usepackage{epstopdf}

\newcount\driver
\newcount\bozza

\font\ottorm=cmr8 scaled\magstep1 
scaled\magstep1                  
scaled\magstep1 \font\msytw=msbm10 scaled\magstep1
 
scaled\magstep1                 \font\indbf=cmbx10 scaled\magstep2

scaled \magstep2

{\count255=\time\divide\count255 by 60
\xdef\hourmin{\number\count255}
        \multiply\count255 by-60\advance\count255 by\time
   \xdef\hourmin{\hourmin:\ifnum\count255<10 0\fi\the\count255}}

\let\a=\alpha \let\b=\beta    \let\g=\gamma     \let\d=\delta     \let\e=\varepsilon
  \let\h=\eta     \let\th=\vartheta      \let\l=\lambda
\let\m=\mu    \let\n=\nu      \let\x=\xi                \let\r=\rho
\let\s=\sigma \let\t=\tau            \let\c=\chi
\let\ps=\psi   \let\o=\omega     
\let\G=\Gamma \let\D=\Delta       \let\L=\Lambda

\def\PP{{\cal P}}\def\EE{{\cal E}}\def\VV{{\cal V}}
\def\HH{{\cal H}}
\def\TT{{\cal T}}\def\BB{{\cal B}}
\def\RR{{\cal R}}\def\LL{{\cal L}}
\def\DD{{\cal D}}\def\GG{{\cal G}}

\def\xx{{\bf x}}
  
\def\yy{{\bf y}}\def\nn{{\bf n}}

\def\tt{{\bf t}}

       \def\oo{{\underline \omega}}
\def\ee{{\underline \varepsilon}}

          \def\BBB{\hbox{\euftw B}}
\def\RRR{\hbox{\msytw R}}

\def\NNN{\hbox{\msytw N}}          
        \def\ZZZ{\hbox{\msytw Z}}

        \def\EE{\hbox{\msytw E}}

\let\==\equiv

\let\io=\infty
\let\0=\noindent

\def\*{{\hfill\break\null\hfill\break}}

\def\tilde#1{{\widetilde #1}}

\def\tende#1{\,\vtop{\ialign{##\crcr\rightarrowfill\crcr
             \noalign{\kern-1pt\nointerlineskip}
             \hskip3.pt${\scriptstyle #1}$\hskip3.pt\crcr}}\,}
\def\otto{\,{\kern-1.truept\leftarrow\kern-5.truept\to\kern-1.truept}\,}

\def\wh#1{\widehat{#1}}
\def\hat#1{\wh{#1}}
\def\sqt[#1]#2{\root #1\of {#2}}

\def\bp{{\bar \ps}}

\def\PP{{\cal P}}\def\EE{{\cal E}}\def\VV{{\cal V}}
\def\HH{{\cal H}}
\def\TT{{\cal T}}\def\BB{{\cal B}}
\def\RR{{\cal R}}\def\LL{{\cal L}}
\def\DD{{\cal D}}\def\GG{{\cal G}}

\def\T#1{{#1_{\kern-3pt\lower7pt\hbox{$\widetilde{}$}}\kern3pt}}
\def\VVV#1{{\underline #1}_{\kern-3pt
\lower7pt\hbox{$\widetilde{}$}}\kern3pt\,}
\def\W#1{#1_{\kern-3pt\lower7.5pt\hbox{$\widetilde{}$}}\kern2pt\,}

\def\indica{\leaders \hbox to 0.5cm{\hss.\hss}\hfill}
\def\guida{\leaders\hbox to 1em{\hss.\hss}\hfill}
\mathchardef\oo= "0521

\def\xx{{\bf x}}
\def\yy{{\bf y}}\def\nn{{\bf n}}

\def\tt{{\bf t}} 
\def\oo{{\underline \omega}}

\def\qed{\raise1pt\hbox{\vrule height5pt width5pt depth0pt}}
 
  \def\bp{{\bar p}} 

\def\indic{\hbox{\raise-2pt \hbox{\indbf 1}}}

\def\RRR{\hbox{\msytw R}} 
 
\def\NNN{\hbox{\msytw N}} 
 \def\ZZZ{\hbox{\msytw Z}}

\def\ins#1#2#3{\vbox to0pt{\kern-#2 \hbox{\kern#1 #3}\vss}\nointerlineskip}

\newdimen\xshift \newdimen\xwidth \newdimen\yshift

\def\insertplot#1#2#3#4#5#6{%
\xwidth=#1pt \xshift=\hsize \advance\xshift by-\xwidth \divide\xshift by 2%
\begin{figure}[ht]
\vspace{#2pt} \hspace{\xshift}
\begin{minipage}{#1pt}
#3 \ifnum\driver=1 \griglia=#6
\ifnum\griglia=1 \openout13=griglia.ps \write13{gsave .2
setlinewidth} \write13{0 10 #1 {dup 0 moveto #2 lineto } for}
\write13{0 10 #2 {dup 0 exch moveto #1 exch lineto } for}
\write13{stroke} \write13{.5 setlinewidth} \write13{0 50 #1 {dup 0
moveto #2 lineto } for} \write13{0 50 #2 {dup 0 exch moveto #1
exch lineto } for} \write13{stroke grestore} \closeout13
\includegraphics{griglia.ps} \fi
\includegraphics{#4.ps}\fi%
\ifnum\driver=2 \fi
\end{minipage}
\caption{#5}
\end{figure}
}

\newdimen\shift \shift=-1.5truecm
\def\lb#1{%
\ifnum\bozza=1
\label{#1}\rlap{\hbox{\hskip\shift$\scriptstyle#1$}}
\else\label{#1} \fi}

\def\be{\begin{equation}}
\def\ee{\end{equation}}
\def\bea{\begin{eqnarray}}\def\eea{\end{eqnarray}}
\def\bean{\begin{eqnarray*}}\def\eean{\end{eqnarray*}}
\def\bfr{\begin{flushright}}\def\efr{\end{flushright}}
\def\bc{\begin{center}}\def\ec{\end{center}}
\def\bal{\begin{align}}\def\eal{\end{align}}
\def\ba#1{\begin{array}{#1}} \def\ea{\end{array}}
\def\bd{\begin{description}}\def\ed{\end{description}}
\def\bv{\begin{verbatim}}\def\ev{\end{verbatim}}
\def\nn{\nonumber}
\def\Halmos{\hfill\vrule height10pt width4pt depth2pt \par\hbox to \hsize{}}
\def\pref#1{(\ref{#1})}

\def\ins#1#2#3{\vbox to0pt{\kern-#2 \hbox{\kern#1 #3}\vss}\nointerlineskip}

\newdimen\xshift \newdimen\xwidth \newdimen\yshift
\newcount\griglia

\def\insertplot#1#2#3#4#5#6{%
\xwidth=#1pt \xshift=\hsize \advance\xshift by-\xwidth \divide\xshift by 2%
\begin{figure}[ht]
\vspace{#2pt} \hspace{\xshift}
\begin{minipage}{#1pt}
#3 \ifnum\driver=1 \griglia=#6
\ifnum\griglia=1 \openout13=griglia.ps \write13{gsave .2
setlinewidth} \write13{0 10 #1 {dup 0 moveto #2 lineto } for}
\write13{0 10 #2 {dup 0 exch moveto #1 exch lineto } for}
\write13{stroke} \write13{.5 setlinewidth} \write13{0 50 #1 {dup 0
moveto #2 lineto } for} \write13{0 50 #2 {dup 0 exch moveto #1
exch lineto } for} \write13{stroke grestore} \closeout13
\includegraphics{griglia.ps} \fi
\includegraphics{#4.ps}\fi%
\ifnum\driver=2 \fi
\end{minipage}
\caption{#5}
\end{figure}
}

\newdimen\shift \shift=-1.5truecm
\def\lb#1{%
\label{#1}\rlap{\hbox{\hskip\shift$\scriptstyle#1$}}
\else\label{#1} \fi}

\def\be{\begin{equation}}
\def\ee{\end{equation}}
\def\bea{\begin{eqnarray}}\def\eea{\end{eqnarray}}
\def\bean{\begin{eqnarray*}}\def\eean{\end{eqnarray*}}
\def\bfr{\begin{flushright}}\def\efr{\end{flushright}}
\def\bc{\begin{center}}\def\ec{\end{center}}
\def\bal{\begin{align}}\def\eal{\end{align}}
\def\ba#1{\begin{array}{#1}} \def\ea{\end{array}}
\def\bd{\begin{description}}\def\ed{\end{description}}
\def\bv{\begin{verbatim}}\def\ev{\end{verbatim}}
\def\nn{\nonumber}
\def\Halmos{\hfill\vrule height10pt width4pt depth2pt \par\hbox to \hsize{}}
\def\pref#1{(\ref{#1})}

\usepackage{amsmath}
\usepackage{amsfonts}
\usepackage{amssymb}
\usepackage{epstopdf}

\font\msytw=msbm9 scaled\magstep1 
scaled\magstep1

\let\a=\alpha \let\b=\beta  \let\g=\gamma  \let\d=\delta
\let\e=\varepsilon
  \let\h=\eta   \let\th=\theta  \let\l=\lambda
\let\m=\mu    \let\n=\nu    \let\x=\xi         \let\r=\rho
\let\s=\sigma \let\t=\tau    \let\c=\chi
\let\ps=\Psi   \let\o=\omega
\let\G=\Gamma \let\D=\Delta  \let\L=\Lambda

\def\EE{{\cal E}} \def\VV{{\cal V}}
 
\def\TT{{\cal T}} \def\BBB{{\cal B}}
\def\RR{{\cal R}}\def\LL{{\cal L}}  
\def\DD{{\cal D}}\def\GG{{\cal G}}

 \def\xx{{\bf x}} \def\yy{{\bf y}} 

\def\PP{{\bf P}}

\def\nn{\nonumber}

\def\RRR{\hbox{\msytw R}} 

\def\NNN{\hbox{\msytw N}} 
 \def\ZZZ{\hbox{\msytw Z}}

\def\\{\hfill\break}
\def\={:=}
\let\io=\infty
\let\0=\noindent

\def\tende#1{\,\vtop{\ialign{##\crcr\rightarrowfill\crcr\noalign{\kern-1pt
    \nointerlineskip} \hskip3.pt${\scriptstyle #1}$\hskip3.pt\crcr}}\,}
\def\otto{\,{\kern-1.truept\leftarrow\kern-5.truept\to\kern-1.truept}\,}

\def\wh{\widehat}
\def\to{\rightarrow}

\def\qed{\hfill\raise1pt\hbox{\vrule height5pt width5pt depth0pt}}

\def\be{\begin{equation}}
\def\ee{\end{equation}}
\def\bp{\begin{pmatrix}}
\def\ep{\end{pmatrix}}
\def\bea{\begin{eqnarray}}
\def\eea{\end{eqnarray}}
\def\nn{\nonumber}
\def\pref#1{(\ref{#1})}

\def\lb{\label}

\newtheorem{lemma}{Lemma}[section]
\newtheorem{theorem}{Theorem}[section]

\begin{document}

\title{Localization in interacting
fermionic chains with quasi-random disorder}

\author{Vieri Mastropietro}

\address{
Universit\'a di Milano, Via C. Saldini 50, 20133, Milano, Italy
}

\begin{abstract}
We consider a system of fermions 
with a quasi-random almost-Mathieu disorder interacting through a many-body short range potential.
We establish exponential decay of the zero temperature correlations, indicating localization
of the interacting ground state, for weak hopping and interaction and almost everywhere
in the frequency and phase; this extends the analysis in
\cite{M} to chemical potentials outside spectral gaps.
The proof is based on Renormalization Group and is inspired by techniques developed to deal with KAM Lindstedt series. 
\end{abstract}


\maketitle

\renewcommand{\thesection}{\arabic{section}}

\section{Introduction and main results}

\subsection{Introduction}

It is due to Anderson \cite{A} the discovery that 
disorder can produce {\it localization} of independent quantum  particles, 
consisting in the exponential decay from some point of the eigenfunctions of the one-body Schroedinger operator. The mathematical understanding of Anderson localization
required the development of powerful techniques
and it was finally rigorously established in the case of random 
\cite{loc0}, \cite{loc1} and quasi-random (or quasi-periodic)
disorder \cite{AA},\cite{S}, \cite{FS},\cite{Ia}. 

A natural 
question is what happens to localization in presence of a many-body interaction, which is always present in real systems. The interplay of disorder and interaction is believed to have deep consequences on the ground state low temperature properties
\cite{FA}, \cite{GT}, \cite{GGG} and
in the non equilibrium  dynamics, like lack of thermalization and memory of initial state
\cite{Ba}, \cite{PH1},\cite{PH4},\cite{H4},\cite{HL}. Mathematical results on localization for interacting systems are still very few \cite{loc4},\cite{M} as the breaking of the 
single particle description makes the problem genuinely infinite dimensional.

In this paper we consider a system of  fermions on a one dimensional lattice with a quasi-random disorder described by a quasi-periodic almost-Mathieu potential $\phi_x=u\cos 2\pi(\o x+\th)$, $\o$ irrational,
and interacting via a short range potential with coupling $U$. Such model is known as the {\it interacting
Aubry-Andr\'e} model \cite{H4},\cite{B} or the {\it Heisenberg quasi-periodic spin chain},
and it has been recently experimentally realized in cold atoms experiments \cite{B}. 

In the absence of interaction the $N$-particle eigenstates can be constructed from the single particle
eigenstates of the Schroedinger energy operator with {\it almost-Mathieu} potential, for which a rather detailed mathematical knowledge exists;
in particular such system shows
a metal-insulator transition, with an Anderson localized insulating phase with
strong disorder and 
a metallic extended phase at weak disorder, similar to what happens in a random three dimensional situation. The {\it exponential} decay of the
single particle
eigenstates of the {\it almost-Mathieu} operator, almost everywhere in $\o,\th$, was proved in \cite{S} and \cite{FS} ,
for $\e$ small enough, $\e$ being the hopping, and later up to
$\e/u$ equal to ${1\over 2}$  in \cite{Ia}. In the opposite regime
$\e/u>{1\over 2}$ the almost Mathieu has {\it extended} states \cite{DS},\cite{P},\cite{E},\cite{BLT},\cite{MP}; 
in particular in \cite{DS} a Diophantine condition is assumed on the phase excluding 
values close to $2\th=\o k$, $k$ integer, corresponding to gaps \cite{MP}. In both regimes and 
for all irrationals the spectrum is a Cantor set \cite{Av}. The non interacting Aubry-Andr\'e model has ground state correlations with a power law decay for large ${\e\over u}$ \cite{BGM}, 
even in presence of interaction \cite{M0},
and an exponential decay for small ${\e\over u}$ \cite{GM}.

The construction of the eigenvectors of the $N$-body Schroedinger equation with almost-Mathieu potential 
and interaction seems at the moment out of reach, especially for infinite $N$; the eigenfunctions
cannot be written as product of eigenfunctions of the single particle operator
and the problem is genuinely infinite dimensional. Information
on the localization of the interacting ground state can be however
obtained by the properties of the zero temperature grand-canonical truncated correlations of local operators, whose exponential decay 
with the distance is a sign
of persistence of localization. We use a technique introduced in \cite{M}
based on a combination of
constructive renormalization Group methods for fermions, see for instance \cite{MM} , with KAM techniques for Lindstedt series \cite{G},\cite{GM1}. 

Our main results can be informally stated as follow.
\vskip.3cm
{\it Almost everywhere in $\o,\th$, for small ${\e\over u}$, ${U\over u}$ 
the zero temperature grand canonical infinite volume truncated correlations  of local operators
decays exponentially for large distances}
\vskip.3cm
The almost everywhere condition in $\o,\th$ is necessary even in the single particle case \cite{Ia};
in particular we assume, as usual,
a Diophantine property for the frequency $\o$ and for the phase  $\th$. In \cite{M} localization was proved assuming that 
$2\th/\o$ integer, corresponding to a choice of the chemical potential in one of the infinitely many gaps; here
we extend such result to a full measure set of phases, where no gap is present.
The result  is in agreement with the qualitative phase diagram obtained by numerical simulations in
\cite{H4} , in which many-body localization is found in the $(U/u,\e/u)$ plane from the origin up to an almost  linear curve
intersecting the points $((U/u)^*,0)$ and $(0,1/2)$, with $(U/u)^*$ of order $1$. Our result establishes localization only for the ground state, but it possible that the method we use can be applied to prove localization of every eigenfunctions of the interacting Schroedinger almost-Mathieu equation.

\subsection{The model}

If  $\L$ is a one dimensional
lattice $\L=\{x\in \mathbb{Z}
, -L/2\le x\le L/2\}$, $L$ even, we
introduce fermionic creation and annihilation operators  $a^+_{x},a^-_{x}$, $x\in\L$
on the Fock space verifying 
$\{a^{+}_x,a^{-}_y\}=\d_{x,y}$, $\{a^{+}_x,a^{+}_y\}=\{a^{-}_x,a^{-}_y\}=0$. The Fock space Hamiltonian is
\be H=-\e(\sum_{x=-{L/2}}^{L/2-1} a^+_{x+ 1} a_{x}+ \sum_{-L/2+1 }^{L/2}a^+_{x-1} a^-_{x} )+\sum_{x=-{L/2}}^{L/2} \phi_x  a^+_{x} a^-_{x}+U \sum_{x,y=-{L/2}}^{L/2}
v(x-y) a^+_{x} a^-_{x} a^+_{y}
a^-_{y}\label{1.1}\ee
with $v(x-y)=\d_{y-x,1}+\d_{x-y,1}$,
and $\phi_x=u \cos (2\pi (\o x+\th))$, $\o$ irrational. {\it We will choose $u=1$ for definiteness}.
If
$a^\pm_\xx=e^{(H-\m N) x_0} a^\pm_x e^{-(H-\m N) x_0}$, $\xx=(x,x_0)$,
$N=\sum_x a^+_x a^-_x$ and $\m$ the chemical potential, 
the Grand-Canonical imaginary time 2-point correlation is
\be
<{\bf T} a^-_{\xx} a^+_{\yy}>|_T={{\it Tr} e^{-\b(H-\m N)}{\bf T}\{a^-_{\xx} a^+_{\yy}
\}\over {\it Tr} e^{-\b(H-\m N)}}\label{asso1}
\ee
where ${\bf T}$ is the time-order product, $T$ denotes truncation and $\m$ is the chemical potential.
In the $\e=U=0$ the spectrum is given by $\sum_x\phi_x n_x$ with $n_x=0,1$ and
the correlations are given by 
the Wick rule in terms of the fermionic 2-point function
 $<{\bf T} a^-_\xx a^+_\yy>|_{U=\e=0}=g(\xx,\yy)$ with
\be
g(\xx,\yy)=\d_{x,y}{1\over\b}\sum_{k_0={2\pi\over\b}(n_0+{1\over 2})}
{e^{-ik_0(x_0-y_0)} 
\over -i k_0+\cos2\pi(\o x+\th) -\m}=\d_{x,y}\bar g(x,x_0-y_0)
\label{prop}
\ee
If
$\m=\cos 2\pi(\o\hat x+\th)$, $\hat x\in\L$ the occupation number, defined as $\bar g(x,0^-)$, is at zero temperature
$\chi(\cos2\pi(\o x+\th) \le\m)$, 
that is
the ground state is obtained by filling all the one particle states with energy $\cos 2\pi(\o x+\th)$ up to the level $\cos 2\pi(\o\hat x+\th)$. 

In the grand canonical ensamble the value of the chemical potential corresponding to a fixed density is a function of the interaction; therefore, if we want to fix the density, what is the more physically natural procedure,
one has to properly choose the chemical potential 
as a function of the interaction. As the 2-point function is singular in correspondence of the chemical potential, this means that the location of the singularity of the 2-point correlation moves varying the interaction; this of course causes problems in a perturbative analysis, resulting in a lack of convergence of a naive power series expansion. It is therefore convenient, both for physical and technical reason, to write the chemical potential 
as a function of the interaction, and to tune it so that the 
singularity in the free or interacting case are the same; this corresponds to fix the density to the same value in the free or interacting case. We therefore write $\m=\cos 2\pi (\o \hat x+\th)+\n$ and we choose properly the counterterm $\n$ as a function of $\e,U$.

The starting point of the Renormalization Group analysis is the representation of the correlations \pref{asso1}
in terms of {\it Grassmann integrals}.  Let $M\in\NNN$ and $\bar\c(t)$ a smooth compact
support function that is $1$ for $t\le 1$ and $0$ for $t\ge \g$, with 
$\g>1$. 
Let $\DD_{\b}=D_{\b}\cap\{k_0\,:\,\bar\c(\g^{-M}|k_0|)>0\}$,
where $D_\b=\{k_0={2\pi\over\b}(n_0+{1\over 2}), n_0\in \ZZZ\}$.
If $x_0-y_0\not= n\b$, we can write
\be\lb{limN} 
g(\xx,\yy) = \lim_{M\to \io} \d_{x,y}\frac1{\b} \sum_{k_0\in \DD_\b}\bar\chi(\g^{-M}|k_0|)
{e^{-ik_0(x_0-y_0)} 
\over -i k_0+\cos2\pi(\o x+\th) -\m}
\equiv \lim_{M\to \io} g^{(\le M)}(\xx,\yy) \ee
Because of the jump discontinuities, $g^{(\le M)}(\xx,\yy)$ is not absolutely
convergent but 
is pointwise convergent
and the limit is given by $g(\xx,\yy)$
at the continuity points, while at the discontinuities it is given by
the mean of the right and left limits. 
If
$\BBB_{\b,L}=\{\L\otimes\DD_\b\}$,   
we consider the
Grassmann algebra generated by the Grassmannian variables
$\{\psi^\pm_{x,k_0}\}_{ x, k_0 \in
\BBB_{\b,L}}$ and a Grassmann
integration $\int
\big[\prod_{x,k_0\in\BBB_{\b,L}}
d\psi_{x,k_0}^- d\psi_{x,k_0}^+\big]$ defined as
the linear operator on the Grassmann algebra such that, given a
monomial $Q(\psi^-, \psi^+)$ in the variables
$\psi^\pm_{x,k_0}$, its action on $Q(\psi^-,
\psi^+)$ is $0$ except in the case $Q(\psi^-,\psi^+)=\prod_{x,k_0\in\BBB_{\b,L}}
\psi^-_{x,k_0}
\psi^+_{x,k_0}$, up to a permutation of the variables. In
this case the value of the integral is determined, by using the
anticommuting properties of the variables, by the condition
\be \int
\Big[\prod_{x,k_0\in\BBB_{\b,L}}
d\psi_{x,k_0}^+
d\psi_{x,k_0}^-\Big]\prod_{x,k_0\in\BBB_{\b,L}}
\psi^-_{x,k_0}
\psi^+_{x,k_0}=1\label{2.1}\ee
We define also Grassmanian field as 
$\psi^\pm_\xx={1\over \b}\sum_{k_0\in \BB_{\b,L}} e^{\pm i k_0  x_0}\psi^\pm_{x,k_0}$ 
with $x_0=m_0 {\b\over \g^M}$ and $m_0\in (0,1,...,\g^M-1)$.
The "Gaussian Grassmann measure"  (also called integration) is defined as
\be
P(d\psi)=[\prod_{x,k_0\in\BBB_{\b,L}}\b
d\psi_{x,k_0}^- d\psi_{x,k_0}^+  
\hat g^{(\le M)}(x,k_0)]\exp\{ -{1\over\b}\sum_{x,k_0} (\hat g^{(\le M)}(x,k_0))^{-1}\psi^+_{x,k_0}\psi^-_{x,k_0}   \} \label{ch}
\ee
with
\be
\hat g^{(\le M)}(x,k_0)={\bar\chi(\g^{-M}|k_0|)
\over -i k_0+\cos2\pi(\o x+\th) -\cos 2\pi(\o \hat x+\th)}
\ee
%
%
We introduce the generating functional $W(\phi)$ defined in terms of the following Grassmann integral  (Dirichelet boundary conditions are imposed)
\be
e^{W(\phi)}=\int P(d\psi)e^{-\VV(\psi)- \BBB(\psi,\phi)}\label{GI}
\ee
with
\bea
&&\VV(\psi)=U\int d\xx \sum_{\a=\pm}
\psi^{+}_{\xx}\psi^{-}_{\xx} \psi^{+}_{\xx+\a{\bf e}_1}\psi^{-}_{\xx+\a{\bf e_1}}
+\e \int d\xx ( t^1_x
\psi^{+}_{\xx+{\bf e_1}}\psi^{-}_{\xx}
+t^2_x\psi^{+}_{\xx-{\bf e_1}}\psi^{-}_{\xx})\nn\\
&& +\n\int d\xx\label{VM}
 \psi^{+}_{\xx}\psi^{-}_{\xx}+\int d\xx  U \n_C(x)
 \psi^{+}_{\xx}\psi^{-}_{\xx}
\eea
where $\int d\xx=\sum_{x\in\L}\int_{-{\b\over 2}}^{\b\over 2} dx_0$,
$t^{1}_{L/2}=t^2_{-L/2}=0$ and $t^1_x=t^2_x=1$ otherwise and
$\n_c(x)=U (\tilde\n_C(x+1)+\tilde\n_C(x-1))$ with
$\tilde\n_C(x)={1\over 2} 
[\bar g(x,0^+)-\bar g(x,0^-)]
$.
Finally
\be
\BBB(\psi,\phi) =   
\int d\xx (\phi^+_{\xx} \psi^-_{\xx} + \psi^+_{\xx}
\phi^-_{\xx})
\ee
%
The 2-point function is given by
\be
S_{2}^{L,\b}(\xx,\yy)
={\partial^2\over\partial\phi^+_{\xx}\partial\phi^{-}_{\yy}}
W|_{0}
\label{asso}
\ee
It is easy to check, see \S 1.C of \cite{M}, that the expansions in $\e, U,\n$ of
\pref{asso1} and of \pref{asso} coincide in the limit $M\to\io$; note in particular the role of the last term of
\pref{VM} taking into account the fact $g(\xx,\yy)$ and $\lim_{M\to \io} g^{(\le M)}(\xx,\yy)$
coincide everywhere except at coinciding points.

\subsection{Main results}

Our main result is the following. 

\begin{theorem} Let us consider the 2-point function $S_{2}^{L,\b}(\xx,\yy)$ \pref{asso} 
with $\m=\cos 2\pi (\o \hat x+\th)$, $\hat x\in \L$, $\hat x , \th$ non vanishing and 
assume that, for some $C_0,\t>1$
\be || \o x||\ge C_0
|x|^{-\t}, \quad ||\o x\pm 2 \th||
\ge C_0 |x|^{-\t}\quad
\forall x\in\ZZZ/\{0\}
\label{d}\ee with $||.||$ is the norm on the one dimensional torus of period $1$.
There exists an $\e_0$ such that, for $|\e|,|U|\le \e_0$ ($u=1$),it is possible to choose a continuous function $\n=\n(\e,U)$ so that the limit $\lim_{\b\to\io}\lim_{L\to\io}\lim_{M\to\io}S_{2}^{L,\b}(\xx,\yy)=S_{2}(\xx,\yy)$  exists and
for any $N\in\NNN$
\be|S_2(\xx,\yy)|\le C e^{- \x|x-y|} \log(1+\min(|x||y|))^\t   {1\over 1+(\D|x_0-y_0)|)^N}\label{fon}\ee
with $\D=(1+\min(|x|,|y|))^{-\t}$, $\x=|\log(\max(|\e|,|U|))|$.
\end{theorem}
\vskip.8cm
The theorem says that the 
ground state correlation decays exponentially for large distances
provided that the hopping $\e/u$ and the interaction $U/u$ are small and for a full measure set of frequencies $\o$ and phases $\th$. The result confirms the phase diagram suggested by numerical experiments 
 \cite{PH4} and says that Anderson localization persists in presence of interaction, at least in the ground state. 
The chemical potential $\m$ is chosen of the form $\m=\cos2\pi(\o \hat x+\th)+\n$, $\hat x\in\NNN$, and the counterterm $\n$ is chosen to fix the density to an $U,\e$-independent value.
The Diophantine condition on the frequency (the first of \pref{d})
is the one usually assumed for proving the localization in the almost Mathieu equation, see
for instance 
\cite{FS}; the second condition in \pref{d} (similar to the one considered for instance in \cite{DS})  
excludes values around integer values of 
${2\th\over \o}$ integer, corresponding to one of the infinitely many gaps in the spectrum.
The values ${2\th\over \o}$ integer
were previously considered in \cite{M} and it was proved that  exponential decay holds and
\pref{fon} is true  
with $\D$ replaced by the gap size.
The above theorem can be equivalently stated fixing the phase $\th$ and varying the chemical potential; 
if we choose $\th=0$ and $\m=\cos 2\pi\o\bar x$, $\bar x\in\RRR$, than the theorem says that the two point function decays exponentially for large distances 
 if $\bar x$ verify a Diophantine condition
$||\o x\pm 2 \o 
\bar x||
\ge C_1 |x|^{-\t}$, $x\not=0$, or if $\bar x$ is half-integer; the first case corresponds to the chemical potential outside gaps while in the second the chemical potential is in the middle of a gap.
The theorem was announced in \cite{M1}.

\subsection{Feynman Graphs expansion and small divisors}

Before starting the proof of Theorem 1.1 it is useful to figure out the main difficulties of the problem, related to the presence of small divisors.
Let us consider the {\it effective potential} defined by
\be
e^{-V(\phi)}=\int P(d\psi)e^{-\VV(\psi+\phi)}\label{GII}
\ee
with $\VV(\psi)$ given by \pref{VM}. We can write
\be V(\phi)=-\log \int P(d\psi) \, e^{-\VV(\psi+\phi)}=\sum_{n=0}^\io {1\over n!}
\EE^T(-\VV;n)\label{fdf12}
\ee
where $\EE^T$ are the {\it fermionic truncated expectations}, that is, if $X(\psi+\phi)$ is a monomial
\be
\EE^T(X:n)={\partial^n\over\partial \a^n}\log \int P(d\psi) e^{\a X(\phi+\ps) })|_{\a=0}
\ee
By evaluating the truncated expectations by the Wick rule, $V(\phi)$ can be written as sum over {\it Feynman graphs}. 

\insertplot{950}{120}
{\ins{170pt}{90pt}{$\xx\pm{\bf e}_1$}
\ins{110pt}{90pt}{$\xx\pm{\bf e}_1$}
\ins{110pt}{20pt}{$\xx$}
\ins{180pt}{20pt}{$\xx$}
\ins{243pt}{30pt}{$\xx$}
\ins{283pt}{30pt}{$\xx+{\bf e}_1$}
\ins{343pt}{30pt}{$\xx$}
\ins{383pt}{30pt}{$\xx$}
\ins{370pt}{10pt}{$\n$}
\ins{270pt}{10pt}{$\e$}
\ins{150pt}{10pt}{$U$}
\ins{460pt}{11pt}{$\n_C(x)$}}%
{verticiT333}
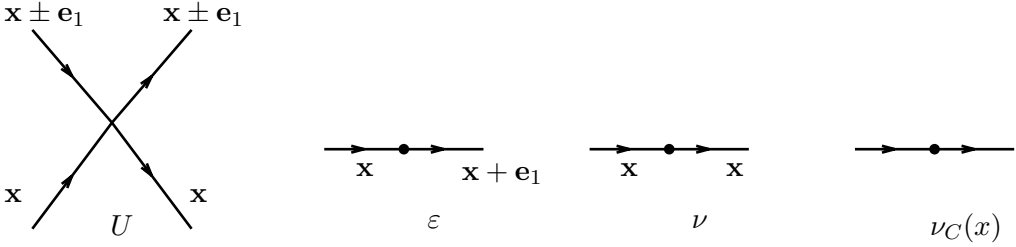
{\label{n9} Graphical representation of the four terms in $\VV(\psi)$ eq.\pref{VM}
}{0}

Each graph 
is obtained taking $n$ elements represented as in Fig.1 
and joining (contracting) 
the lines with consistent orientation so that all the $n$ vertices are connected.
Calling $\ell$ the contracted lines of the graph, if $\GG_n$ is the set of all possible  
Feynman graphs of order $n$, for any graph $G\in \GG_n$ 
we can associate a {\it value} ${\rm Val}(G)$; for instance the graphs not involving the last term in \pref{VM} have the value, if $n=n_U+n_\e+n_\n$ and $\int d\xx=\int dx_0\sum_x$ 
\be{\rm Val}(G)=(-1)^\pi U^{n_U} \e^{n_\e}\n^{n_\n}
\int d\xx_1...\int d\xx_n
\prod_\ell g(\xx_\ell,\yy_\ell)\prod_{i\in A(G)} \phi^{\s_i}_{\xx_i}
\label{salsa}
\ee
where $A(G)$ is the set of indices of the non contracted lines, 
$\ell$ are the contracted lines of the graph and $\xx_\ell,\yy_\ell$ the coordinates
at the edge of the line, and $(-1)^\pi$ is the sign associated to the graph. 
With the above definitions
\be
V(\phi)=
\sum_{n=0}^\io\sum_{G\in \GG_n}{\rm Val}(G)\label{ex556}
\ee
In the non interacting case $U=0$ the only possible graphs are chain graphs; an
example is in Fig. 2.

\insertplot{890}{100}
{\ins{220pt}{25pt}{$\e$}
\ins{260pt}{25pt}{$\e$}
\ins{300pt}{25pt}{$\e$}
\ins{360pt}{25pt}{$\e$}}%
{verticiT444}
{\label{n9} A graph with $n_\e=4$, $n_U=n_\n=0$
}{0}

The value is 
\bea
&&\e^{\n_\r}\n^{n_\n}\int \prod_{1=1}^n d\xx_i \phi_{\xx_1}[\prod_{i=1}^{n}
\d_{x_i+\a_i,x_{i+1}} \bar g( x_i+\a_i,  x_{0,i}-x_{0,i+1})]
\phi_{\xx_{n+1}}=\nn\\
&&e^{\n_\r}\n^{n_\n}
\sum_{x_1}\int d x_{0,1}...dx_{0,n}
\phi_{\xx_1}
\phi_{x_1+\sum_{i\le n} \a_i,x_{0,n}}\prod_{i=1}^n \bar g(x_1+\sum_{k\le i}\a_k,x_{0,i+1}-x_{0,i})\nn
\eea
which can be rewritten as
\be
\e^{\n_\e}\n^{n_\n}
\sum_{x_1}\int dk_0 
\hat \phi_{x_1,k_0}[\prod_{k=1}^n\hat g(x_1+\sum_{i\le k} \a_i,k_0)]
\hat\phi_{x_1+\sum_{i\le k} \a_i,k_{0}}=\e^{\n_\e}\n^{n_\n}
\sum_{x_1}\int dk_0 H(k_0,x_1)
\label{gigi}
\ee
%
%
%
In order to bound  $H(k_0,x_1)$
we note that, 
as the frequency $\o$ is irrational, $(\o x)_{mod. 1}$
fills densely the interval $(-1/2,1/2]$ so that the denominator $\phi_x -\m$
can be {\it arbitrarily small}.
Let us introduce $
\bar x_+=\hat x\quad \bar x_-=-\hat x-2\th/\o
$.
%
%
%
If we set $x=x'+\bar x_\r$, $\r=\pm$,
for small $(\o x')_{\rm mod. 1}$
then $\cos 2\pi (\o(x'+\bar x_\r)+\th)
-\cos (2\pi( \o\hat  x+\th))=
\r v_0 (\o x')_{\rm mod. 1}
+r_{\r,x'}$  with 
$r_{\r, x'}=O(((\o x')_{\rm mod. 1})^2)$, $v_0=\sin 2\pi(\o\hat x+\th)$ ,
so that, for small $(\o x')_{\rm mod. 1}$
\be
\hat g(x'+\bar x_\r, k_0)\sim {1\over -i k_0\pm v_0  (\o x')_{\rm mod. 1}}
\label{l3}
\ee
Note that, for $x\not=\r\hat x$ 
\be
||\o x'||=||\o (x-\r\hat x)+2\d_{\r,-1}\th||\ge C |x-\r\hat x|^{-\t}
\ee
by \pref{d}. Therefore the sum of all the chain graphs of order $n$ is bounded by
$\e^n C^n ||\hat x|+|n||^{\t n}$, a bound which does not imply convergence.

In the case of the interacting theory the graphs are much more  complex and loops are present; an example is Fig. 3 
\insertplot{890}{220}
{\ins{160pt}{120pt}{$U$}
\ins{260pt}{120pt}{$\e$}
\ins{220pt}{120pt}{$\e$}
\ins{300pt}{120pt}{$\e$}
\ins{310pt}{120pt}{$U$}
\ins{230pt}{190pt}{$\e$}}%
{verticiT555}
{\label{n9} A graph with $n_U=2,n_\e=4$
}{0}
whose value is the following
\bea
&&\e^{4}U^2 \sum_x \int dx_{0,1}...dx_{0,6} \phi_{x} \bar g(x;x_{0,1}-x_{0,2}) g(x+1,x_{0,2}-x_{0,3})
\bar g(x;x_{0,3}-x_{0,4})\\&&\bar g(x+1;x_{0,4}-x_{0,5})
\bar g(x+1;x_{0,1}-x_{0,5})\bar g(x+1;x_{0,1}-x_{0,6})\bar g(x+2;x_{0,6}-x_{0,5})\phi_{x+2,x_{5,0}}\nn\label{hj}
\eea
Note that, after summing over the coordinates and exploiting the kronecker deltas
of the propagators connecting the vertices, only a single sum over $x$ remains.
Again each graph does not admit a bound which allow us to sum over $n$; in addition there is the problem that 
the number of graphs with loops is $O(n!^2)$ (the number of chain graphs is $O(n!)$ instead).
Note that the small divisor problem of the non interacting theory 
is very similar to the one appearing in KAM theory; for instance the Lindstedt series can be represented
in terms of graphs with no loops very similar to \pref{gigi}, see \cite{G}.
On the contrary the appearance of graphs with loops plagued by small divisors like \pref{hj} is the peculiar feature of localization in a many body theory.

\section{Proof of Theorem 1.1}

\subsection{Multiscale integration and Renormalization Group analysis}

We start describing the integration of the generating function in the case $\phi=0$ (the partition function).
We introduce a function
$\c_h(t,k_0) \in C^{\io}(\mathbb{T} \times  \mathbb{R})$,
such that $\c_h(t,k_0) = \c_h(-t,-k_0)$ and $\c_h(t,k_0) = 1$, if
$\sqrt{k_0^2+v_0^2 ||t||^2_1}\le a \g^{h-1}$
and  $\c_h(t,k_0) = 0$ if $\sqrt{k_0^2+v_0^2 ||t||^2_1}\ge a \g^h$
with $a$ and $\g>1$ suitable constants. We define
$
\bar x_+=\hat x\quad \bar x_-=-\hat x-2\th/\o
$
and
we choose $a$ so that
the supports of $\c_0(\o(x-\hat x_+),k_0)$ and $\c_0(\o(x-\hat x_-),k_0)$
are disjoint;
we also define $\chi^{(1)} (\o x,k_0) = 1- \c_0(\o(x-\bar x_+),k_0) -
\c_0(\o(x-\bar x_-),k_0)$.
For reasons which will appear clear below, see Lemma 2.4,
we choose $\g> 2^{1\over\t}$. 
We  can write then
\be
g(\xx,\yy)=g^{(1)}(\xx,\yy)+g^{(\le 0)}(\xx,\yy)
\ee
and
\be
g^{(\le 0)}(\xx,\yy)=\sum_{\r=\pm} g_\r^{(\le 0)}(\xx,\yy)
\ee
where, for $M$ large enough
\bea 
&&g^{(1)}(\xx,\yy)={\d_{x,y}\over \b}\sum_{k_0\in
D_{\b}} \chi^{(1)} (\o x,k_0)\bar\chi(\g^{-M}|k_0|)
{e^{-i k_0(x_0-y_0)}\over -i k_0+\cos2\pi(\o x+\th) -\cos 2\pi(\o \hat x+\th)
} \nn\\
&&g^{(\le 0)}_\r(\xx,\yy)={\d_{x,y}\over \b}\sum_{k_0\in
D_{\b}}\c_0(\o(x-\bar x_\r),k_0) 
{e^{-i k_0(x_0-y_0)}\over -i k_0+\cos2\pi(\o x+\th) -\cos 2\pi(\o \hat x+\th)} 
\eea
We use the following property; if $P_g(d\psi)$ is a Gaussian Grassmann integration with propagator $g$
and $g=g_1+g_2$, then $P_g(d\psi)=P_{g_1}(d\psi_1)P_{g_2}(d\psi_2)$,
in the sense that for every polynomial $f$
\be 
\int P_g(d\psi) f(\psi)=\int P_{g_1}(d\psi_1)\int P_{g_2}(d\psi_2)
f(\psi_1+\psi_2)\;.\label{fonfo1}
\ee
By using such property
\be e^{W(0)}=\int P(d\psi) e^{-\VV(\psi)} =
\int P(d\psi^{(\le 0)})\int P(d\psi^{(1)}) e^{-\VV(\psi^{(\le 0)}+\psi^{(1)})}\label{V0} \ee
where $P(d\psi^{(1)})$ and $P(d\psi^{(\le 0})$ are gaussian Grassmann integrations with propagators
respectively $g^{(1)}(\xx,\yy)$ and $g^{(\le 0)}(\xx,\yy)$ and $\psi^{(1)}$ and $\psi^{(\le 0)}$ 
are independent Grassmann variables. We can write 
\be
\int P(d\psi^{(1)}) e^{-\VV(\psi^{(\le 0)}+\psi^{(1)})}=e^{\sum_{n=0}^\io{(-1)^n\over n!}\EE^T_{1}(\VV:n)}\equiv e^{-\b L E_0-\VV^{(0)}(\psi^{(\le 0)})}\label{cxc}
\ee
where $\EE^T_1$ is the fermionic truncated expectation with respect to $P(d\psi^{(1)})$. 
%
%
By the above definition
\be
\VV^{(0)}=\sum_{n=1}^\io \sum_{x_1}\int dx_{0,1}....\sum_{x_n}\int dx_{0,n}
W_{n}^{(h)}(\xx_1,...,\xx_n) [\prod_{i=1}^n \psi^{(\e_i)(\le 0)}_{\xx'_i,\r_i}]
\ee
with $\xx=\xx'+\bar\xx_\r$, 
$
\bar\xx_\r=(\bar x_\r,0)$
and $E_0$ is a constant; moreover
\be e^{W(0)}=e^{-\b L E_0}\int P(d\psi^{(\le 0)})e^{-\VV^{(0)}(\psi^{(\le 0})} \ee
Note that the kernel $W_{n}^{(h)}
(\xx_1,...,\xx_n)$ contains in general Kronecker or Dirac
deltas, and we define the $L_1$ norm as they would be positive
functions, {\it e.g.}  if $W(\xx_1,\xx_2,..\xx_n)=\d(\sum_j \h_j \xx_j) \bar W(\xx_1,..,\xx_n)$ then $|W|_{L_1}=
\int d\xx_1..d\xx_n \d(\sum_j \h_j \xx_j) |\bar W(\xx_1,..,\xx_n)|$.
It was proved in Lemma 2.1 \cite{M} 
that
the constant $E_{0}$ and the kernels
$W^{(0)}_{n}$ are given by power series in $U,\e,\n$
convergent for $|U|,|\e|, |\n|\le \e_0$, for $\e_0$ small enough and
independent of $\b,L$. They satisfy the following bounds:
\be |W^{(0)}_{n}|_{L_1}  \le  L\b C^{n}
\e_0^{k_{n}}\;,\lb{2.17}\ee
for some constant $C>0$ and $k_{n}=\max\{1,n-1\}$. Moreover the limit $M\to\io$ exists and is reached uniformly.
 
We describe the integration of $\psi^{(\le 0)}$ inductively.
Assume that we have integrated the fields $\psi^{(0)}...\psi^{(h+1)}$
obtaining
\be  e^{-\b L E_0}\int P(d\psi^{(\le 0)})e^{-\VV^{(0)}(\psi^{(\le 0)})}=e^{-\b L E_h} \int P(d\psi^{(\le h)})e^{-\VV^{(h)}(\psi^{(\le h)})}\label{gr1} \ee
where $P(d\psi^{(\le h)})$ is the gaussian grassman integration with propagator, $\r=\pm $
\be
g^{(\le h)}_{\r}(\xx',\yy')=\d_{x',y'}\bar g^{(\le h)}_{\r}(x',x_0-y_0)
\ee
with, if $x=x'+\bar x_\r$
\be
g^{(\le h)}_{\r}(x',x_0-y_0)= \int dk_0 e^{-i k_0 (x_0-y_0)}\chi_h(\o x',k_0){1\over -i
k_0+ v_0 \r (\o x')_{\rm mod. 1}+r_{\r,x'}}
\label{prop1}
\ee
and the corresponding fields are denoted by $\psi^{(\e,\le h)}_{\xx',\r}$. 
The effective potential 
$\VV^{(h)}$ can be written as sum of terms (see \S 2 B below )
of the form
\be \sum_{x'_1}\int dx_{0,1}....\int dx_{0,n}
H_{n;\r_1,..,\r_n}^{(h)} (x'_1; x_{0,1},.,x_{0,n}) [\prod_{i=1}^n
\psi^{\e_i(\le h)}_{\xx'_i, \r_i}]\label{cd}\ee
and $x'_i$ are functions of $x_1$. There is an important constraint on the $\r$ indices; if $x'_i=x'_j$ then $\r_i=\r_j$. This follows from the second of \pref{d}, implying that ${2\th\over\o}\not\in\ZZZ/\{0\}$; 
indeed as $x_i-x_j=M\in\ZZZ$ and $x'_i=x'_j$ then
$(\bar x_{\r_i}-\bar x_{\r_j})+M=
0$, so that
$\r_i=\r_j$ as 
$\bar x_+=\hat x$ and $\bar x_-=-\hat x-2\th/\o$
and
$\hat x\in \ZZZ$.

We call {\it resonances} the contribution to $\VV^{(h)}$ of the form
\pref{cd} 
such that $x'_i=x'_1\equiv x'$ for any $i=1,..,n$; 
in the resonances $\r'_i=\r'_1$ for any $i=1,..,n$.
%
%

In order to perform the integration
of the field $\psi^{(h)}$ we have to split the effective potential as $\VV^{(h)}=\LL\VV^{(h)}+\RR\VV^{(h)}$
where $\RR=1-\LL$ and $\RR$ 
is defined in the following way.
\begin{enumerate}
\item If $n=2$ then $\RR=1$ if \pref{cd} is non resonant, while if \pref{cd} is resonant
\bea &&\RR \sum_{x'}\int dx_{0,1} dx_{0,2}
H_{2;\r,\r}^{(h)} (x'; x_{0,1},x_{0,2}) 
\psi^{+(\le h)}_{x',x_{0,1}, \r}\psi^{-(\le h)}_{x',x_{0,2}, \r}\label{r11}\\
&&=\sum_{x'}\int dx_{0,1} dx_{0,2}\{ H_{2;\r,\r}^{(h)} (x'; x_{0,1},x_{0,2}) 
\psi^{+(\le h)}_{x',x_{0,1}, \r}\psi^{-(\le h)}_{x',x_{0,2}, \r}
- H_{2;\r,\r}^{(h)} (0; x_{0,1},x_{0,2}) 
\psi^{+(\le h)}_{x',x_{0,1}, \r}\psi^{-(\le h)}_{x',x_{0,1}, \r} \}\nn\eea
%
%
%
\item If $n=4$ $\RR=1$  if \pref{cd} is non resonant, while if \pref{cd} is resonant
\bea
&&\RR \sum_{x'}\int \prod_{i=1}^4 dx_{0,i}  
H_{4;\r,\r,\r,\r}^{(h)} (x'; x_{0,1},x_{0,2},x_{0,3},x_{0,4}) 
\psi^{+(\le h)}_{x',x_{0,1}, \r}\psi^{+(\le h)}_{x',x_{0,2}, \r}
\psi^{-(\le h)}_{x',x_{0,3}, \r}\psi^{-(\le h)}_{x',x_{0,4}, \r}=\label{r22}\\
&&\sum_{x'}\int  \prod_{i=1}^4 dx_{0,i}
H_{4;\r,\r,\r,\r}^{(h)} (x'; x_{0,1},x_{0,2},x_{0,3},x_{0,4}) 
D^{+(\le h)}_{x',x_{0,1},x_{0,2} \r}\psi^{+(\le h)}_{x',x_{0,2}, \r}
D^{-(\le h)}_{x',x_{0,3},x_{0,4} \r}\psi^{-(\le h)}_{x',x_{0,4}, \r}\nn
\eea
where
%
%
\be
D^{\pm(\le h)}_{x', x_{0,1},x_{0,2},\r}=\psi^{\pm(\le h)}_{x', x_{0,1},\r}-\psi^{\pm(\le h)}_{x', x_{0,2},\r}
\ee
%
That is,
the $\RR$ operation simply consists in replacing the fields $\psi^{\pm(\le h)}_{x',x_{0,i}, \r}\psi^{\pm(\le h)}_{x',x_{0,j}, \r}$ with $D^{\pm(\le h)}_{x',x_{0,i},x_{0,j}, \r}\psi^{\pm(\le h)}_{x',x_{0,j}, \r}$
\item If $n\ge 6$, the $\RR$ operation consists in replacing any monomial of fields with the same $x,\e$ in 
\pref{cd},
that is
$\psi^{\e(\le h)}_{x',x_{0,1},\r}\prod_i  \psi^{\e(\le h)}_{x',x_{0,i},\r}$, with 
\be \psi^{\e(\le h)}_{x',x_{0,1},\r}\prod_i  D^{\e(\le h)}_{x',x_{0,1},x_{0,i},\r}\label{r4}\ee.
\end{enumerate}
By the above definitions
\be \LL \VV^{(h)}= \g^h \n_h \sum_{\r}\sum_{x'}\int dx_0 \psi^{+(\le
h)}_{\xx',\r}
\psi^{-(\le h)}_{\xx',\r}\label{zak}
\ee
%
%
The $\n_h$ coefficients are {\it independent} from $\r$ and real, 
as
\pref{GI} is invariant under parity $x\to -x$, $\a\to-\a$ (in the limit $L\to\io$), and this
implies invariance under the transformation
$\psi^{\pm (h)}_{x_0,x',\r}
\to \psi^{\pm(h)}_{x_0,-x',-\r}$; therefore, if $\e=\pm $
\be
\sum_{\r, x'} 
\int dx_0 dy_0  H^{(h)}_{2,\r}(x',x_0,y_0)\psi^{+(\le h)}_{x',x_0,\r}
\psi^{+(\le h)}_{x', x_0,\r}=\sum_{\r, x'} 
\int dx_0 dy_0  H^{(h)}_{2,-\r,}(-x',x_0,y_0)\psi^{+(\le h)}_{x',x_0,\r}
\psi^{+(\le h)}_{x',x_0,\r}
\ee
so that the independence from $\r$ of $\n_h$ follows.
Moreover $(g^{(k)})^*(x,k_0)=g^{(k)}(x,-k_0)$ so that $(\hat H^{(h)}_{2,\r}(x',k_0))^*=
\hat H^{(h)}_{2,\r}(x',-k_0)$, and this implies reality. In writing \pref{zak} we have also used that 
$\psi^\pm_{x', x_{0,1},\r}\psi^\pm_{x',x_{0,1},\r}=0$, so that there is no contribution from non bilinear terms .
With the above definitions we finally write \pref{gr1} as
\be
\int  P(d\psi^{(\le h-1)})\int P(d\psi^{(h)}
)e^{-\LL\VV^{(h)}-\RR\VV^{(h)}}=e^{-\b L \tilde E_h}
\int P(d\psi^{(\le h-1)}) e^{-\VV^{(h-1)}(\psi^{(\le h-1)})}\label{sss}
\ee
where $P(d\psi^{(\le h-1)})$ have
propagator $g^{(\le h-1)}$ coinciding with 
\pref{prop1} with $h-1$ replacing $h$, and 
$P(d\psi^{(h)})$ has propagator $g^{(h)}$ coinciding with 
$g^{(\le h-1)}$ with $\chi_{h-1}$ replaced by $f_h=\chi_h-\chi_{h-1}$,
with
$f_h$ a smooth compact support function vanishing for 
$c_1\g^{h-1}\le \sqrt{k_0^2+v_0^2 ||\o x'||_1^2}\le c_2\g^{h+1}$, for a suitable constants $c_1,c_2$.
%
%
From the r.h.s. of \pref{sss}, the procedure can be iterated.  
The single scale propagator $g^{(h)}$ verifies the following bound, for any integer $N$  and a
suitable constant $C_N$
\be
|\bar g^{(h)}_\r(x',x_0-y_0)| \le {C_N\over 1+(\g^h|x_0-y_0|)^N}\label{za}
\ee
which can be easily obtained integrating by parts.

The above procedure allows to write the $W(0)$
\pref{V0} in terms of an expansion in the {\it running coupling constants} $\n_k,$ with $k\le 0$; as it is clear from the above construction, they verify a recursive equation of the form
\be
\n_{h-1}=\g \n_{h}+ \b^{(h)}(\n_h,..
\n_0;\e;U)\label{beta}
\ee
We will describe more explicitly such expansion in the following section.


\subsection{Tree expansion}

The effective potential $\VV^{(h)}$ can be written as sum over Gallavotti {\it trees}, 
defined in the following way. Let us consider the family of all trees which can be constructed
by joining a point $r$, the {\it root}, with an ordered set of
$n\ge 1$ points, the {\it endpoints} of the {\it unlabeled
tree}, so that $r$ is not a branching point. $n$ will be
called the {\it order} of the unlabeled tree and the branching
points will be called the {\it non trivial vertices}. The
unlabeled trees are partially ordered from the root to the
endpoints in the natural way; we shall use the symbol $<$ to
denote the partial order, and their number is bounded by $4^n$.
\insertplot{400}{195}
{\ins{50pt}{80pt}{$v_0$}
\ins{50pt}{-5pt}{$h$}
\ins{95pt}{-5pt}{$h_{v'}$}
\ins{120pt}{-5pt}{$h_{v}$}
\ins{100pt}{90pt}{$v'$}
\ins{115pt}{95pt}{$v$}
\ins{220pt}{-5pt}{$0$}\ins{240pt}{-5pt}{$1$}\ins{260pt}{-5pt}{$2$}}
{fig51}
{\label{h2q} 
A tree $\t\in\TT_{h,n}$ with its scale labels.}
{0}
We shall also consider the set
$\TT_{h,n}$ of the {\it labeled trees} with $n$ endpoints (to be
called simply trees in the following); they are defined by
associating some labels with the unlabeled trees.
In particular,  we associate a label $h\le 0$ with the root. Moreover, we
introduce a family of vertical lines, labeled by an integer taking
values in $[h,2]$, and we represent any tree $\t\in\TT_{h,n}$
so that, if $v$ is an endpoint or a non trivial vertex, it is
contained in a vertical line with index $h_v>h$, to be called the
{\it scale} of $v$, while the root $r$ is on the line with index
$h$. In general, the tree will intersect the vertical lines in set
of points different from the root, the endpoints and the branching
points; these points will be called {\it trivial vertices}. 
Every vertex $v$ of a tree will be
associated to its scale label $h_v$, defined, as above, as the
label of the vertical line whom $v$ belongs to. Note that, if
$v_1$ and $v_2$ are two vertices and $v_1<v_2$, then
$h_{v_1}<h_{v_2}$.

There is only one vertex immediately following the root, which will be
denoted $v_0$; its scale is $h+1$.  Given a vertex $v$ of $\t\in\TT_{h,n}$ that is not an
endpoint, we can consider the subtrees of $\t$ with root $v$,
which correspond to the connected components of the restriction of
$\t$ to the vertices $w\ge v$; the number of endpoint of these
subtrees will be called $N_v$. If a subtree with root $v$ contains
only $v$ and one endpoint on scale $h_v+1$, it will be called a
{\it trivial subtree}. With 
each endpoint $v$ of scale $h_v\le 1$ we associate $\LL\VV^{(h_v-1)}$,
and there is the constrain that
$h_v=h_{v'}+1$, if $v'$ is the non trivial vertex immediately preceding
it or $v_0$; to the end-points of scale $h_v=2$ are
associated one of the terms
contributing to  $\VV$ and there is not such a constrain.
The set of field labels
associated with the endpoint $v$ will be called $I_v$;
if $v$ is not an endpoint, we shall call $I_v$ the
set of field labels associated with the endpoints following the
vertex $v$.
Finally with each trivial or non trivial vertex $v>v_0$, $h_v\le 0$, which is not an endpoint, we
associate the $\RR=1-\LL$ operator, acting on the corresponding kernel.

If $h\le -1$ the effective potential can be written in the following way:
\be
\VV^{(h)}(\psi^{(\le h)}) + L\b E_{h+1}=
\sum_{n=1}^\io\sum_{\t\in\TT_{h,n}}
V^{(h)}(\t,\psi^{(\le h)})
\ee
where, if $v_0$ is the first vertex of $\t$ and $\t_1,..,\t_s$ ($s=s_{v_0}$)
are the subtrees of $\t$ with root $v_0$,\\
$V^{(h)}(\t,\psi^{(\le h)})$ is defined inductively by the relation, if $s>1$
\be
V^{(h)}(\t,\psi^{(\le h)})=
{(-1)^{s+1}\over s!} \EE^T_{h+1}[\bar
V^{(h+1)}(\t_1,\psi^{(\le h+1)});..; \bar
V^{(h+1)}(\t_{s},\psi^{(\le h+1)})]\label{3.33}
\ee
where $\bar V^{(h+1)}(\t_i,\psi^{(\le h+1)})$:
\begin{enumerate}
\item 
it is equal to $\RR\VV^{(h+1)}(\t_i,\psi^{(\le h+1)})$, with $\RR$ given by \pref{r11},\pref{r22},\pref{r4} if
the subtree $\t_i$ is non trivial;
\item if $\t_i$ is trivial, it is equal to $\LL\VV^{(h+1)}$.
\end{enumerate}
Starting from the above inductive definition, the effective potential can be written in a more explicit way.
We associate with any vertex $v$ of the tree a subset $P_v$ of
$I_v$, the {\it external fields} of $v$, and the set $\xx_v$ of
all space-time points associated with one of the end-points
following $v$. The subsets $P_v$ must satisfy various constraints.
First of all, $|P_v|\ge 2$, if $v>v_0$; moreover, if $v$ is not an
endpoint and $v_1,\ldots,v_{S_v}$ are the $S_v\ge 1$ vertices
immediately following it, then $P_v \subseteq \cup_i P_{v_i}$; if
$v$ is an endpoint, $P_v=I_v$. If $v$ is not an endpoint, we shall
denote by $Q_{v_i}$ the intersection of $P_v$ and $P_{v_i}$; this
definition implies that $P_v=\cup_i Q_{v_i}$. The union ${\cal
I}_v$ of the subsets $P_{v_i}\setminus Q_{v_i}$ is, by definition,
the set of the {\it internal fields} of $v$, and is non empty if
$S_v>1$. Given $\t\in\TT_{h,n}$, there are many possible choices
of the subsets $P_v$, $v\in\t$, compatible with all the
constraints. We shall denote ${\cal P}_\t$ the family of all these
choices and ${\bf P}$ the elements of ${\cal P}_\t$.
With these definitions, we can rewrite $\VV^{(h)}(\t,\psi^{(\le
h)})$ as
\be \VV^{(h)}(\t,\psi^{(\le h)})=\sum_{{\bf P}\in{\cal
P}_\t} \VV^{(h)}(\t,{\bf P})\quad \bar\VV^{(h)}(\t,{\bf P})=\int d\xx_{v_0} \widetilde\psi^{(\le
h)}(P_{v_0}) K_{\t,{\bf P}}^{(h+1)}(\xx_{v_0})\;,\lb{2.43a}\ee
where $K_{\t,{\bf P}}^{(h+1)}(\xx_{v_0})$ is defined inductively and
$
\widetilde\psi^{(h_v)}(P_{v})=\prod_{f\in P_v}\psi^{\e(f)(h_v)}_{\xx'(f),\r(f)}$.
%
%

Given a tree $\t$ and ${\bf P}\in {\cal P}_\t$ , we shall define the
$\chi$-vertices are the vertices $v$ of $\t$ ,
such that
${\cal I}_v$ (the union of the subsets $P_{v_i}\setminus Q_{v_i}$ 
defined before \pref{2.43a}, that is the set of lines contracted in $v$) is non empty; note that $|V_\chi|$ is smaller than $4n$. We call $\bar v'$ is the first vertex $\in V_\chi$ following $v$.
The tree structure 
provides an arrangement of endpoints into a hierarchy of {\it clusters}, see Fig.5.  Given a cluster with scale $h_v$, one can imagine that the fields 
$\widetilde\psi^{(h_v)}(P_{v_1}\setminus Q_{v_1})$,..,$
\widetilde\psi^{(h_v)}(P_{v_{S_v}}\setminus
Q_{v_{S_v}})$ are external to the $S_v$ inner clusters, and the $\EE^T_{h_v}$ operation contracts them in pairs.
\insertplot{340}{160}
{\ins{125pt}{140pt}{$1$}
\ins{125pt}{100pt}{$2$}
\ins{125pt}{80pt}{$3$}
\ins{125pt}{60pt}{$4$}
\ins{125pt}{20pt}{$5$}
\ins{145pt}{80pt}{$\Longleftrightarrow$}
\ins{193pt}{75pt}{$1$}
\ins{237pt}{75pt}{$2$}
\ins{253pt}{75pt}{$3$}
\ins{261pt}{75pt}{$4$}
\ins{291pt}{75pt}{$5$}}
{fig10}
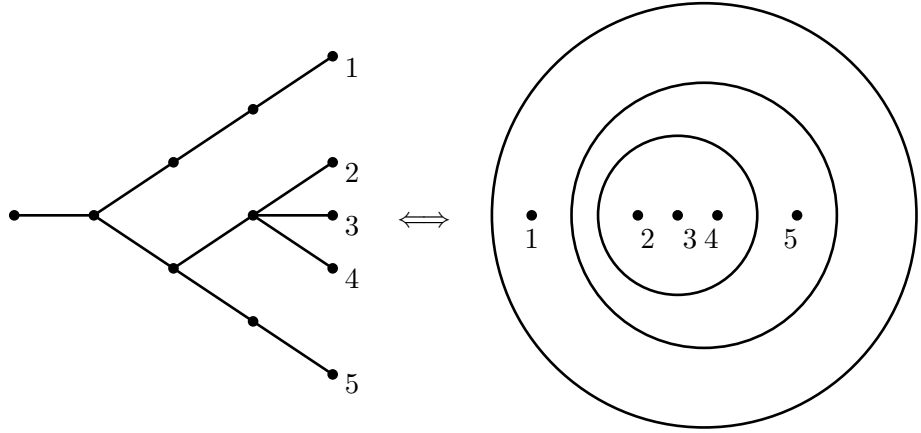
{\label{n10} A tree of order 5 and the corresponding
clusters. Only the vertices $v\in V_\chi$ are represented.}{0}

In order to get the final form of our expansion, we need a
convenient representation for the truncated expectation. Let us put $P_i\=P_{v_i}\setminus
Q_{v_i}$; moreover we order in an arbitrary way the sets
$P_{v_i}^\pm\=\{f\in P_{v_i},\e(f)=\pm\}$, we call $f_{ij}^\pm$ their
elements and we define $\xx^{(i)}=\cup_{f\in P_{i}^-}\xx(f)
$,
$\yy^{(i)}=\cup_{f\in P_{i}^+}\yy(f)$, $\xx_{ij}=\xx(f^-_{ij})
$,
$\yy_{ij}=\xx(f^+_{ij})$. A couple
$l\=(f^-_{ij},f^+_{i'j'})\=(f^-_l,f^+_l)$ will be called a line
joining the fields with labels $f^-_{ij},f^+_{i'j'}$. Then, we use
the {\it Brydges-Battle-Federbush} formula saying that
,  if $S_v>1$,
\be \EE^T_{h_v}(\tilde\psi^{(h_v)}(P_{i}),\cdots,
\tilde\psi^{(h_v)}(P_{S_v})
))=\sum_{T_v}\prod_{l\in T_v}
\big[\d_{x_l,y_l}\bar g^{(h_v)}_{\r_l}(x'_l,x_{0,l}-y_{0,l})\big]\, \int dP_{T}({\bf t})\; {\rm
det}\, G^{h_v,T}({\bf t})\;,\label{2.46aa}\ee
where $T_v$ is a set of lines forming an {\it anchored tree graph} between the
clusters of points $\xx^{(i)}\cup\yy^{(i)}$, see Fig.6,
that is $T_v$ is a set of lines,
which becomes a tree graph if one identifies all the points in the same
cluster. Moreover ${\bf t}=\{t_{ii'}\in [0,1], 1\le i,i' \le S_v\}$,
$dP_{T_v}({\bf t})$ is a probability measure with support on a set of ${\bf t}$
such that $t_{ii'}={\bf u}_i\cdot{\bf u}_{i'}$ for some family of vectors
${\bf u}_i\in \RRR^{S_v}$ of unit norm.
\be G^{h_v,T}_{ij,i'j'}=t_{ii'}
\d_{x_{ij},y_{i'j'}} \bar g^{(h_v)}_{\r_{ij}}(x_{ij}, x_{0,ij}-y_{0,i'j'})\;,
\label{2.48}\ee
with $(f^-_{ij}, f^+_{i'j'})$ not belonging to $T_v$.
\insertplot{1490}{220}
{}%
{verticiT888}
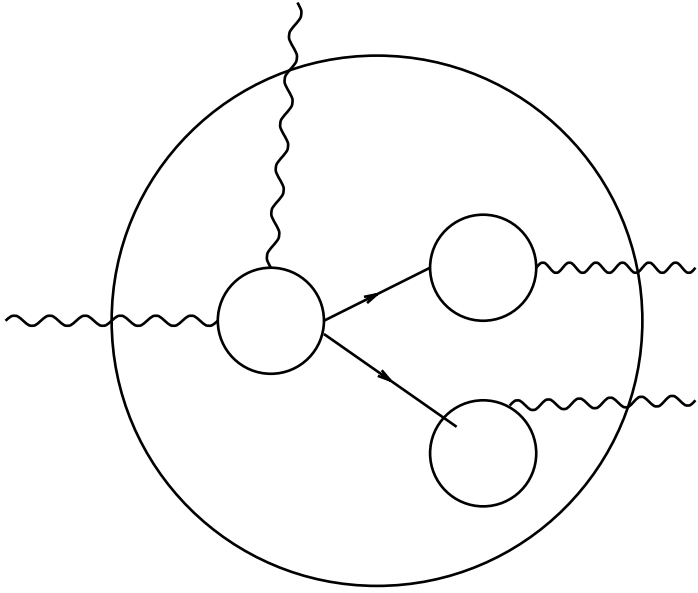
{\label{n9} A symbolic representation of a contribution to \pref{2.46aa}; the solid lines represent the propagators $g^{(h_v)}$ in $T_v$ connecting the $S_v=3$ clusters, represented as circles,
the wiggly lines are the external fields $\tilde\psi(P_v)$; the fields in the determinant are not represented}{0}

We define $\bar T_v=\bigcup_{w\ge v}T_w$ starting
from $T_v$ and attaching to it the trees $T_{v_1},..,T_{v_{S_v}}$ associated to the vertices $v_1,..,v_{S_v}$
following $v$, and repeating
this operation until the end-points are reached.
The tree $\bar T_v$ is composed by 
a set of lines, representing propagators with scale $\ge h_v$, connecting end-points $w$ of the tree $\t$. Note that, contrary to $T_v$, the vertices of $\bar T_v$ are connected with at most four lines. 
By writing the truncated expectations as in \pref{2.46aa} we write $\VV^{(h)}$ as sum  over $T_v$, for any $v$; 
by summing the Kronecker deltas in the propagators belonging to $T_v$ the coordinate $x'$ of the external fields $\tilde\psi(P_v)$ are determined according to the following rule. 
To each 
line coming in or out $w$ is associated a factor $\d_{w}^{i_w}$, where $i_w$
is a label identifying the lines connected to $w$. The vertices $w$ (which correspond to the end-points of $\t$) can be of type $U,\n$ or $\n_h$, and a) $\d^i_w=0$ if $w$ corresponds to a $\n$ or $\n_h$ end-point;
b) $\d_w^i=\pm 1$ if it corresponds to an $\e$ end-point;
c) $\d^i_w=(0, \pm 1)$ if it corresponds to a $U$ end-point.

\insertplot{500}{195}
{\ins{100pt}{110pt}{$w_1$}
\ins{75pt}{100pt}{$w_a$}
\ins{45pt}{80pt}{$w_b$}
\ins{20pt}{80pt}{$w_c$}
\ins{40pt}{40pt}{$w_2$}
}
{fig60}
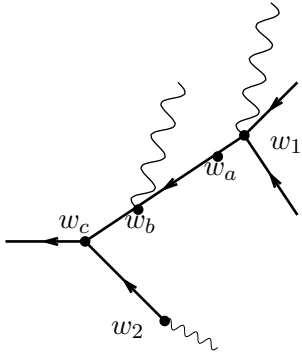
{\label{h2} A tree $\bar T_v$ with attached wiggly lines representing the 
 external lines $P_v$; the lines
represent propagators with scale $\ge h_v$ connecting $w_1,w_a,w_b,w_c,w_2$, representing
the end-points following $v$ in $\t$.  
} {0}

According to the above definitions, consider two
vertices $w_1,w_2$ in $\bar T_v$ such that $x'_{w_1}$ and $x'_{w_2}$ are
coordinates of the external fields, and  let be $c_{w_1,w_2}$ the path (vertices and lines)
in $\bar T_v$ connecting $w_1$ with $w_2$ (in the example in Fig. 7 the path is composed by $w_1,w_a,w_b,w_c,w_2$ and the corresponding lines)
; as the path is a linear tree there is a natural orientation in the vertices, and we we call $i_w$ the label of the line exiting from $w$ in 
$c_{w_1,w_2}$. We call $|c_{w_1,w_2}|$ the number of vertices in $c_{w_1,w_2}$.
The following relation holds
\be x'_{w_1}-x'_{w_2}= (\bar x_{\r_{\ell_{w_2}}}-\bar x_{\r_{\ell_{w_1}}})+ \sum_{w\in c_{w_1,w_2}}
\d_{w}^{i_{w}}\label{fa}
\ee
This implies, in particular, that the coordinates of the external fields $\tilde\psi(P_{v_0})$ are determined once that the choice of a single one of them and of $\t, \bar T_{v_0}$ and ${\bf P}$ is done.
If, using \pref{fa}, the coordinates $x'$ of the fields $\tilde\psi(P_v)$ are the {\it same} we say that $v$ is a 
{\it resonant vertex},
while if the coordinates are different is called {\it non resonant vertex}; the set of resonant vertices in $V_\chi$ is denoted by $H_\chi$ and the set of non-resonant vertices is denoted by $L_\chi$.
If $v_1,\ldots,v_{S_v}$
are the $S_v\ge 1$ vertices following the vertex $v$, we define
\be
S_v=S^L_v+S^H_v+S^2_v\label{bra}
\ee
where $S^L_v$ is the number of {\it non resonant} vertices following $v$,
$S^H_v$  is the number of {\it resonant} vertices following $v$,
while $S_v^2$  is the number of trivial trees with root $v$ associated to end-points.

\subsection{Renormalization}

In order to get the final form of our expansion we need to write more explicitly 
the action of the renormalization operation $\RR$; we can write the r.h.s. of \pref{r11} as
\bea&&
\int dx_{0,1}dx_{0,2}
\{ H_{2;\r,\r}^{(h)} (x'; x_{0,1},x_{0,2})- H_{2;\r,\r}^{(h)} (0; x_{0,1},x_{0,2})\}
\psi^{+(\le h)}_{x',x_{0,1}, \r}\psi^{-(\le h)}_{x',x_{0,2}, \r}+\nn\\
&&
H_{2;\r,\r}^{(h)} (0; x_{0,1},x_{0,2}) 
D^{+(\le h)}_{x',x_{0,1},x_{0,2} \r}\psi^{-(\le h)}_{x',x_{0,1}, \r}\}\label{jj}
\eea
The second term in \pref{jj} consists in replacing the $\psi$ fields with $D$-fields;
the same effect is produced by the $\RR$ operation in \pref{r22}, \pref{r4}.
The propagators associated to the $D$ fields are
\be \bar g^{(h)}(x', x_{0,1}-z_0)-\bar g^{(h)}(x',x_{0,i}-z_0)\label{diff}
\ee
which can be conveniently rewritten as 
\be
(x_{0,1}-x_{0,i}) \int_0^1 dt \partial \bar g^{(h)}
(x',\hat x_{0,1i}(t)-z_0)
\label{df11} \ee
where $\hat x_{0,1i}(t)=x_{0,1}+t(x_{0,i}-x_{0,1})$ is an interpolated point between $x_{0,1}$
and $x_{0,2}$. Note that the "zero" factor $(x_{0,1}-x_{0,i})$, produces an extra $\g^{-k}$ in the bounds and the extra derivative produces an extra $\g^{h}$; the final factor is $\g^{h-k}$.
The difference $H_{2;\r,\r}^{(h)} (x'; x_{0,1},x_{0,2})- H_{2;\r,\r}^{(h)} (0; x_{0,1},x_{0,2})$
in \pref{jj} can be written as a sum of terms in which a propagator $\bar g^{(k)}(x'+y;z_0)$ is replaced by 
\be \bar g^{(k)}(x'+y;z_0)-\bar g^{(k)}(y;z_0)\label{dif0}
\ee which can be rewritten as
\bea
&&\bar g^{(k)}(x'+y, z_0)-g^{(k)}(y,z_0)=\\
&&(\o x' )\int dk_0 e^{-i k_0 z_0}\int_0^1 {\partial\over \partial t \o x'} {f_h(\o y+t \o x',k_0)\over -i k_0+\cos2\pi(\o y+\o\bar x_\r+t\o x'+\th)-\cos 2\pi(\o\bar x_\r+\th)}\nn\label{dif01}
\eea
Note that $(\o x' )\sim \g^{h}$ for the compact support properties of the propagators associated to $\psi^{\le h}$, while the derivative produces an extra $\g^{-k}$; therefore the final effect is again to produce an extra $\g^{h-k}$ factor in the bounds. 

\subsection{Renormalized Graphs expansion}

We can write the truncated expectations in terms of the Wick rule, and this leads to 
a representation of the effective potential in terms of renormalized Feynman graphs 
\be\VV^{(h)}(\psi^{(\le h)})=\sum_{n=1}^\io \sum_{\t\in \TT_{h,n}}\sum_{G\in \GG(\t)}{\rm Val}(G)
\label{esp16}
\ee
where $\GG(\t)$ is the set of renormalized 
Feynman graphs; with respect to the Feynman graph described in \S 1.D, 
each propagator 
carries an index $h_v$, if $v\in V_\chi$ is the minimal cluster containing the propagator, see Fig.8.
\insertplot{880}{210}
{\ins{115pt}{170pt}{$U$}
\ins{115pt}{135pt}{$U$}
\ins{115pt}{100pt}{$U$}
}
{fig9a}
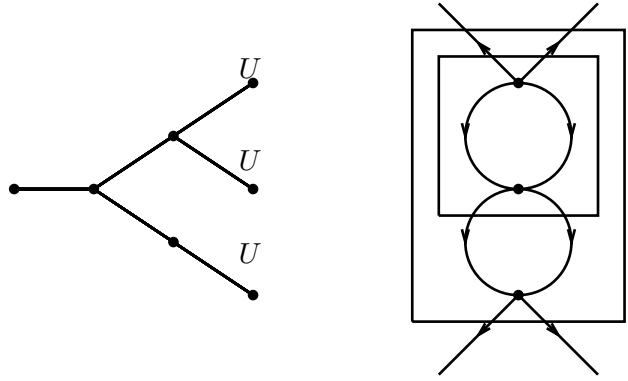
{\label{n11} A tree $\t$ (only the vertices $v\in V_\chi$ are represented), the corresponding clusters, represented as boxes, and a Feynman graph; the propagators have scale $h_{v_1}$ and $h_{v_2}$ respectively. 
}{0}
If we do not take into account the $\RR$ operation, an immediate bound for each Feynman graph is, if $|\n_h|,|U|,|\e|\le \e_0$, and remembering that $S_v$ is the number of clusters contained in the cluster $v$
\be
\e_0^n C^n \prod_{v\in V_\chi} \g^{-(S_v-1)h_v }\label{prob}
\ee
The above estimate is obtained considering a tree of propagators connecting all vertices, bounding
by a constant the propagators not belonging to such tree and by $\g^{-h_v}$ the integrals of each one of the 
$S_v-1$ propagators in the tree connecting the vertices in the cluster $v$.
Note that, as $h_v<0$, the above estimate is unbounded when summed over $h_v$. 
By definition we can rewrite \pref{prob} as
\be
\e_0^n C^n\prod_{v\in V_\chi} \g^{-(S^H_v+S^L_v+S^2_v-1)h_v }\label{prob}
\ee
If we take into account the $\RR$ operation than, by \S 2.C, we get the following bound 
\be
\e_0^n C^n\prod_{v\in V_\chi} \g^{-(S^H_v+S^L_v+S^2_v-1)h_v }\prod_{v\in H_\chi}
\g^{(h_{\bar v'}-h_v)}
\label{prob1}
\ee
where $\bar v'$ is the first vertex $\in V_\chi$ following $v$
and the last factor can compensate the term $\g^{-S^H_v h_v }$, proportional to the number of resonant terms,
 from the first factor in \pref{prob1}; this is indeed the reason why we introduce the $\RR$ operation.
It remains however the term $\g^{-S^L_v h_v }$, proportional to the number of non resonant terms; one cannot define the $\RR$ operation for the non resonant terms, as in that way 
an infinite number of relevant terms with any number of fields is produced
(in absence of the resonance condition the local part is non vanishing), and we could not control their flow.
As we will see in the following section, the contribution of the non resonant terms is controlled using the Diophantine conditions.

\subsection{The non resonant terms}

%
%
%
%

Consider a non resonant vertex $v$ and $x'_{w_1}$ and $x'_{w_2}$ are
coordinates of two external fields, with $x'_{w_1}-x'_{w_2}$ given by \pref{fa}.
The Diophantine conditions imply a relation between the
scale $h_v$ and the number of vertices between $w_2$ and
$w_1$ in $\bar T_v$.

\begin{lemma} Given $\t, {\bf P}, {\bf T}$, let us consider $v\in L_\chi$
and $w_1, w_2$ two vertices in $\bar T_v$, see \pref{fa}, with $x'_{w_1}\not =x'_{w_2}$; then  
\be |c_{w_1,w_2}|\ge A \g^{-h_{\bar  v'}\over \t}\label{h19}
\ee with a suitable constant $A$.
\end{lemma}

{\it Proof.} Note that $||\o x'_{w_i}||_1\le c v_0^{-1} \g^{h_{\bar v'}-1}$, $i=1,2$ by the compact support properties of the propagator; therefore by
using \pref{fa} and the Diophantine condition, if 
\bea
&&2 c v_0^{-1}\g^{h_{\bar v'}}\ge ||(\o x'_{w_1})||+ ||(\o x'_{w_2})||\ge 
||\o(x'_{w_1}-x'_{w_2})||=
\\
&&||(\bar x_{\r_{\ell_{w_2}}}-\bar x_{\r_{\ell_{w_1}}})\o
+
\o\sum_{w\in c_{w_1,w_2}}
\d_{w}^{i_{w}}||
\eea
If $\r_{\ell_{w_2}}=\r_{\ell_{w_1}}$ by the first of \pref{d} we get
\be 2 c v_0^{-1}\g^{h_{\bar v'}}\ge {C_0\over |\sum_{w\in c_{w_1,w_2}}
\d_{w}^{i_{w}}|^{-\t}}\ee. 

If $\r_{\ell_{w_2}}=\e$, $
\r_{\ell_{w_1}}=-\e$, $\e=\pm$ then 
\be
||(\bar x_{\r_{\ell_{w_2}}}-\bar x_{\r_{\ell_{w_1}}})\o
+
\o\sum_{w\in c_{w_1,w_2}}
\d_{w}^{i_{w}}||=||2\e\o\hat x+2\e\th+
\o\sum_{w\in c_{w_1,w_2}}
\d_{w}^{i_{w}}||
\ee
and if 
$\sum_{w\in c_{w_1,w_2}}
\d_{w}^{i_{w}}+2\e\hat x\not= 0$ by the second of \pref{d}
\be
2 c v_0^{-1}\g^{h_{\bar v'}}\ge {C_0\over |2\e\hat x+\sum_{w\in c_{w_1,w_2}}
\d_{w}^{i_{w}}|^{-\t}}\ge  {C_0\over (2|\hat x|+|\sum_{w\in c_{w_1,w_2}}
\d_{w}^{i_{w}}|)^{-\t}}\ge  {C_0\over |\sum_{w\in c_{w_1,w_2}}
\d_{w}^{i_{w}}|^{-\t}}
\ee
Finally if $\sum_{w\in c_{w_1,w_2}}
\d_{w}^{i_{w}}+2\e\hat x= 0$ then 
$c v_0^{-1}\g^{h_{\bar v'}}\ge ||2\th||\ge  ||2\th|| {|2\hat x|^\t\over|\sum_{w\in c_{w_1,w_2}}
\d_{w}^{i_{w}}|^{\t}}$. The fact that
$|\sum_{w\in c_{w_1,w_2}}
\d_{w}^{i_{w}}|\le |c_{w_1,w_2}|$ ends the proof.
\qed \vskip.3cm 

Lemma 2.1 says that there is a relation between the number
of end-points following $v\in L_\chi$ and the scales of the external lines coming out from $v$.
In particular the $U,\e$-endpoints with scale $h_ v=2$ have $|c_{w_1,w_2}|=1$, hence the scale of the first vertex $v\in V_\chi$ preceding the end-point is bounded by a constant.

%
\begin{lemma}
Given $\t, {\bf P}, {\bf T}$
the following inequality holds, for any $0<c<1$
\be c^{n}\le \prod_{v\in L_\chi} c^{ A \g^{-h_{\bar v'}\over \t}2^{h_{ \bar v'}-1}} \label{zza} \ee
\end{lemma}
\vskip.3cm
{\it Proof}. If $v\in V_\chi$ and $N_v=\sum_{i, v^*_i>v} 1$ is
the number of end-points following $v$ in $\t$ then
\be
c^{n} \le
\prod_{v \in V_\chi} c^{N_v 2^{h_{\bar v'}-1}}\label{org}\ee
Indeed we can write
\be
c=\prod_{h=-\io}^0 c^{2^{h-1}}
\ee
Given a tree $\t\in \TT_{h,n}$, we consider an end-point $v^*$ and the path in $\t$ from $v^*$
to the root $v_0$; to each vertex $v\in V_\chi$ in such path with scale $h_v$ we associate a factor 
$c^{2^{h_v-2}}$; repeating such operation for any end-point, the vertices $v$ followed by $N_v$ end-points are in $N_v$ paths, therefore  we can associate to them a factor  $c^{N_v 2^{h_{ v}-2}}$; finally
we use that $c^{2^{h_{ v}-2}}<c^{2^{h_{\bar v'}-2}}$.

Note that if $v$ is non resonant, there exists surely two external fields
with coordinates $x'_1, x'_2$ such that $x'_1\not=x'_2$; note that
\be
N_v\ge |c_{w_1,w_2}|\ge A  \g^{-h_{\bar v'}\over \t}
\ee
therefore, by \pref{org}, \pref{zza} follows,
%
%
\qed.
\vskip.6cm

By combing the above results we get the following final lemma which will play a crucial role in the following.
We choose $\g^{1\over\t}/2\equiv \g^\h>1$; for instance $\g=2^{2\t}$, $\h={1\over 2\t}$.

\begin{lemma}
Given $\t, {\bf P}, {\bf T}$
the following inequality holds
\be
[\prod_{v\in V_\chi} \g^{-4h_v S^L_v}]
[
\prod_{v\in L_\chi}
c^{A \g^{-h_{v'}\over \t}2^{h_{v'}}}]\le 
\bar C^n
\label{zzaa} \ee
with  $\bar C= [{3\over |\log |c|| A}]^3 e^{-3}]$.
\end{lemma}

{\it Proof}
As we assumed $\g^{1\over\t}/2\equiv \g^\h>1$ than, for any $N$
\be
c^{A \g^{-h\over \t}2^{h}}=e^{-|\log c|
A \g^{-\h h}}\le \g^{N \h h} {N\over |\log |c|| A ]^N e^{N}}
\ee
as  $e^{-\a x} x^N\le [{N\over \a}]^N e^{-N}$. Therefore, by choosing $N=4/\h$ we get
\be\prod_{v\in L_\chi} c^{A  \g^{-h_{\bar v'}\over \t}2^{h_{\bar v'}}} \le \bar  C^n\prod_{v\in V_\chi} \g^{4 S^L_{ v} h_{v}}\ee
%
\qed
\vskip.6cm

\subsection{Renormalized expansion}

We write the expansion for the kernels of the effective potential in a way more suitable for the final bounds. Given a
contribution with fixed $\t,{\bf P},T$ to $\VV^{(h)}$, we consider the vertex $v$ in $\t$ with smallest $h_v$ on which the $\RR$ operation acts non trivially. Let us consider first the case
of a resonance with two external lines.  We consider a $D$-field associated  
to one of the external lines, see the second term in \pref{jj}, and we write it as \pref{df11}.
We decompose the zero as $(x_{0,1}-x_{0,j})=\sum_k (x_{0,k}-x_{0,k+1})$, where 
$(x_{0,k}-x_{0,k+1})$ is a zero corresponding to one of the lines $l$ of the tree graph $\bar T_v$.
If the corresponding propagator has scale $h_w$, and if $w>w'>w''...>v$, we
add an index
to one of the external D-fields (if present) of each vertex between
$w$ and $v$ indicating that, in the next iteration, one has not to write the corresponding difference of propagators as \pref{df11} (the contribution of the two terms is written separately). The reason is that one gets in the bounds an extra factor 
$\g^{h_v-h_w}=\g^{h_{w'}-h_w}\g^{h_{w''}-h_{w'}}...$, so that the gain of the $\RR$ operation on the intermediate vertices is already obtained (for more details see  
for instance \S 3 of \cite{BM} in a similar case). Similarly we proceed for the first term in \pref{jj}.
If $v$ is resonant and has 4 external fields, we proceed in the same way; if there is a zero of order $2$ in a propagator in $w$, we add an index to
two of the D-fields (if present) of each vertex between
$w$ and $v$ saying that one has not to write the corresponding difference of propagators as \pref{df11}.
Finally assume that $v$ has more than $6$ external fields (resonant or non resonant);
 we call $\bar\r,\bar\e$ the labels of the external fields whose number is maximal; 
we define this set $m_v$ and
$|m_v|\ge {|P_v|/4}$.
We consider a tree $\bar T_v$ and 
we define a pruning operation
associating to it another tree $\hat T_v$
eliminating from $\bar T_v$ all the trivial vertices $w$ in $\bar T_v$ not associated 
to any external line
with label $\bar \r,\bar\e$, and all the 
subtrees not containing any external line
with label $\bar \r,\bar\e$ (see Fig. 9 for an example), so that there is an external line associated to all end-points. 
\insertplot{700}{195}
{
\ins{80pt}{100pt}{$w_6$}
\ins{140pt}{160pt}{$w_1$}
\ins{125pt}{100pt}{$w_{2}$}
\ins{100pt}{140pt}{$w_3$}
\ins{115pt}{80pt}{$w_4$}
\ins{90pt}{110pt}{$w_5$}
\ins{50pt}{80pt}{$w_7$}
\ins{40pt}{65pt}{$w_8$}
\ins{90pt}{35pt}{$w_9$}
\ins{10pt}{35pt}{$w_{10}$}
\ins{10pt}{65pt}{$w_{11}$}
\ins{40pt}{115pt}{$w_{12}$}
}
{fig61}
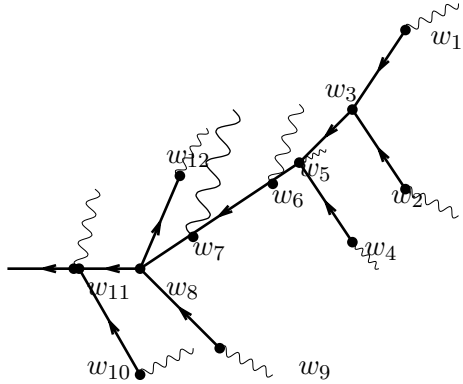
{\label{h2} In the picture the lines represent the propagators with scale $\le h_v$ in $\hat T_v$
and the wiggly lines represent the external lines $P_v$ with label $\bar\r$; note that, by definition of the pruning operation,
all the end-points have associated wiggly lines, contrary to what happens in $\bar T_v$, see Fig. 7.
} {0}
The vertices $w$ of $\hat T_v$
are then only non trivial vertices or trivial vertices with external lines $\bar\r,\bar\e$; all the end-points have associated an external line.
We define a procedure to group in two sets
the fields in $m_v$. We start
considering the end-points $w_a$ immediately followed by vertices $w_b$ with external lines (in the figure  
$w_4,w_{10}$), and we say that the couple of fields in $w_a,w_b$ is of type 1
if $x'_{w_a}=x'_{w_b}$, while it is of type 2 if  $x'_{w_a}\not =x'_{w_b}$.
We now prune tree $\hat T_v$ canceling the end-points $w_a$ already considered 
and the resulting subtrees with no external lines; in the resulting tree we select
an end-point $w_a$ immediately followed by vertices $w_b$, and again such a  couple can be of type 1 or 2. 
We again prune the tree and we continue unless there are no end-points $w$ followed by vertices with wiggly line.
Then in the resulting tree  we consider (if they are present, otherwise the tree is trivial and the procedure ends)
a couple of endpoints followed by the same non trivial vertex (in the picture $w_1,w_2$); we call them $w_a,w_b$ and we proceed exactly as above distinguishing the two kind of couples. We then cancel such end-points $w_a,w_b$
and the subtrees not containing external lines, so that the end-points are associated to external lines;
we consider end-points followed by non trivial vertices with no external lines, and we proceed in the same way. If the resulting tree 
has again end-points with external lines followed by vertices with external lines (in the picture $w_5$), 
we prune such vertices as described above and
we continue in this way so that at the end all except at most one vertex with external lines are considered.
Note that by construction the paths 
$c_{w_a,w_b}$ in $\bar T_v$
do not overlap; 
for instance in Fig.9 the paths are 
$c_{w_{10},w_{11}}$, $c_{w_4,w_5}$, $c_{w_1,w_2}$, $c_{w_5,w_{6}}$, $c_{w_6,w_{7}}$,
$c_{w_{7},w_{12}}, c_{w_9,w_{11}}$. 
Therefore, given a vertex $v$ in the tree $\t$, we have paired all the external fields with index $\bar\r,\bar\e$, whose number is 
$|m_v|\ge {|P_v|/4}$, in couples both with the same $x'$ or with different $x'$.
We say that a field 
is of type $1$ if it belongs to a couple with the same $x'$ and of type $2$ if belongs to a couple
with different $x'$ (if it belongs to 2 couples of different kind, we follow the order of the construction).
The number of fields of type $1$ is $|m^1_v|$ and 
type $2$ is $|m^2_v|$ and $|m_v|=|m^1_v|+|m^2_v|$. 
In a couple of fields with the same $x'$ one is surely a $D$-fields; we then write it as 
\pref{df11} which will produce in the bounds a factor $\g^{(h_{\bar v'}-h_v)}$. Note that 
the zero is decomposed along the path connecting the two fields; as the paths 
$c_{w,w'}$ are non overlapping by construction, the order of such zero is at most $1$; by this fact we get in the bounds a factor $\g^{(h_{\bar v'}-h_v){|m^1_v|\over 2}}$.
Again an index is added to the $D$
fields associated to vertices between the vertex of the zero and $v$, as done above.
On the other hand given $w,w'$ with $x'_w
\not=x'_{w'}$, 
we have
$|c_{w,w''}|\ge B\g^{-h_{\bar v'}/\t}$ by lemma 2.1; moreover 
by Lemma 2.2 we can associate to each $v\in V_\chi$ a factor $c^{N_v 2^{h_{\bar v}-1}}$
with $N_v$ the vertices in $\bar T_v$;
as the paths $c_{w,w'}$ are non overlapping, we get one factor 
$
c^{|c_{w,w'}|2^{h_{v'}}}\le c^{B  \g^{-h_{\bar v'}/\t}  2^{h_{\bar v'}}}\label{f}
$ for each of the couples .
The above procedure is the iterated in the vertices $\hat v$ following $v$ in $\t$, taking into account the indices saying that some of the $D$ fields is not written as \pref{df11}.
%
%

We add an index $\a$ to distinguish the terms generated by this procedure, so that we can write
\be V^{(h)}=\sum_{n=1}^\io \sum_{\t\in \TT_{h,n}} \sum_{T\in {\bf T}}
 \sum_{{\bf P}\in {\cal P}_\t}\sum_{\a\in A_T}
\sum_x \int dx_{0,v_0} H_{\t,{\bf P},T,\a}(x,
x_{0,v_0})\prod_{f\in P_{v_0}}\psi^{(\le h)\e(f)}_{\hat
\xx'(f),\r(f)}
\label{lau}\ee
and
\bea 
&&H_{\t,{\bf
P},T,\a}(x, x_{0,v_0})=K_{\t,{\bf P},T,\a}\prod_{v\ {\rm not}\ {\rm e.p.}}{1\over S_v!}
\int dP_{T}({\bf t})\;
{\rm det}\, \tilde  G^{h_v,T_v}_\a({\bf t}_v)\label{vvvv}\\
&&
\prod_{l\in T_v}\partial^{q_\a(f^+_l)}_{\g^{h_v}x_{0,l}}\partial^{q_\a(f^-_l)}_{\g^{h_v}x_{0,l}}
\partial^{\tilde q_\a(f^+_l)}_{\g^{-h_v}\o x'_{l}}
( \g^{h_l}(x_{0,l}-y_{0,l}))^{b_\a(l)} (\g^{-h_v}(\o x'_l))^{\tilde b_\a(l)}
\bar g_{\r_l}^{(h_v)}(x'_l ; x_{0,l}-y_{0,l}))
\big|\Bigg]
\nn \eea
where
${\bf T}$ is the set of the tree graphs on $\xx_{v_0}$, obtained by
putting together an anchored tree graph $T_v$ for each non trivial vertex $v$,
$A_T$ is a
set of indices which allows to distinguish the different terms produced by
the non trivial $\RR$ operations and the iterative decomposition of the zeros
$G_\a^{h_v,T_v}(\tt_v)$ has elements
\be G^{h_v,T_v}_{\a,ij,i'j'}=t_{v,i,i'}
\d_{x_{ij},y_{i'j'}}(\o x_{ij})^{\tilde q_\a(f^+_{ij})}
\partial^{q_\a(f^+_{ij})}_{\g^h x_{0ij}}\partial^{q_\a(f^-_{ij})}
_{\g^h x_{0ij}} g^{(h)}(x_{ij}, x_{0,ij}-y_{0,i'j'})
\ee
The indices $q_\a,\tilde q_\a,b_\a, \tilde b_\a\in (0,3)$ are such that, by construction and for $c<1$
\be
|K_{\t,{\bf P},T,\a}|\le c^{-n}
\prod_{v\in V_\chi} \g^{(\a_v+\b_v)(h_{\bar v'}-h_{v})}\g^{-\a |P_v|}\label{35}
\ee
with $\bar v'$ the first vertex belonging to $V_\chi$ following $v$ in $\t$ and, by construction
\begin{enumerate}
\item if $v$ is resonant then $\a_v=1$;
\item If $v$ is resonant and $|P_v|\ge 4$
then $\b_v=1$
\end{enumerate}

The factor $\prod_{v\in V_\chi} \g^{(\a_v+\b_v)(h_{\bar v'}-h_{v})}$ is obtained by the action of $\RR$ on the resonant term; the factor $\g^{-\a |P_v|}$ is obtained, as discussed above \pref{lau}, by the action of $\RR$ on the terms with more than $6$ lines and by Lemma 2.2
\be
\prod_{v\in V_\chi}
c^{\g^{-h_{\bar v'}/\t}  2^{h_{\bar v'}}|m^2_v|
}
\prod_{v\in V_\chi} \g^{{1\over 82}|m^{1}_v|(h_{\bar v'}-h_v)}
\le \prod_{v\in V_\chi}\g^{-\a |P_v|}\label{34}
\ee
%
Note that
\be
\prod_{v\in V_\chi} \g^{(\a_v+\b_v)(h_{\bar v'}-h_{v})}=\prod_{v\in H_\chi}\g^{(1+\b_v)(h_{\bar v'}-h_{v})}\label{fonfon}
\ee

Regarding the flow equation  for $\n_h$ we get by construction
\be
\n_{h-1}=\g\n_h+\g^{-h}\sum_{n\ge 2}\sum_{\t\in \TT_{h,n}} \sum_{T\in {\bf T}}
 \sum_{{\bf P}\in {\cal P}_\t}\sum_{\a\in A_T}
\int dx_{0,v_0} H_{\t,{\bf P},T}(0,
x_{0,v_0})\label{sper}
\ee
Note that on the first vertex of the trees $v_0$ the $\LL$ operation acts; therefore, as $\LL\RR=0$,
necessarily $v_0\in V_\chi$.

\subsection{Bounds for the effective potential}

In this section we get a bound for the kernels of the effective potential defined in \pref{lau}.

\begin{lemma} If $n=n_\n+n_U+\n_\e$ the following bound holds
\be 
{1\over\b L} \sum_{\t\in {\cal T}_{h,n}} \sum_{T\in {\bf
T}}\sum_{{\bf P}\in \PP_\t}\sum_x \int dx_{0,v_0}
|H_{\t,{\bf P},T,\a}(x,
x_{0,v_0})|\le C^n  \g^{h_{v_0}}
(\sup_{k\ge h} |\n_k||)^{n_\n}|U|^{n_U}|\e|^{n_\e}\label{bra1}
\ee
where $C$ is a suitable constant.
%
\end{lemma}
\vskip.3cm {\it Proof} 
We start from \pref{vvvv} and, in order to
bound the matrix $\tilde G^{h,T}_{ij,i'j'}$, we introduce an Hilbert
space $\HH=\ell^2\otimes\RRR^s\otimes L^2(\RRR^1)$ so that
\be
\tilde G^{h,T}_{ij,i'j'}=
\Big({\bf v}_{x_{ij}}\otimes {\bf u}_{i}\otimes
A(x_{0, ij}-,x_{ij})\;,\ {\bf v}_{y_{i',j'}}\otimes
{\bf u}_{i'}
\otimes B(y_{0,i'j'}-,x_{ij})\Big)
\label{as}
\;,\ee
where ${\bf v}\in \RRR^L$ are unit vectors such that $({\bf v}_i,{\bf v}_j)=\d_{ij}$,
${\bf u}\in \RRR^{s}$ are unit vectors $(u_i,u_{i})=t_{ii'}$, and $A,B$
are vectors in the Hilbert space with scalar product \be (A,B)=\int dz_0 A(x',x_0-z_0)B^*(x',z_0-y_0)\ee
given by
\be A(x',x_0-z_0)={1\over\b}\sum_{k_0} e^{-i k_0
(x_0-z_0)}\sqrt{f_h(\o x', k_0)}\nn\ee
\be
B(x',y_0-z_0)={1\over \b}\sum_{k_0} {e^{-ik_0( y_0-z_0)}\sqrt{f_h(\o x', k_0)}
\over -i k_0+\cos 2\pi (\o x'+\bar x_\r+\th)-\cos 2\pi (\bar x_\r+\th)}\nn
\label{2.48b}\ee
Moreover
\be ||A_h||^2=\int dz_0 |A_h(x',z_0)|^2\le C\g^{-3h}\;,\quad\quad
||B_h||^2\le C \g^{3h}\;,\lb{B.5}\ee
for a suitable constant $C$. Therefore
by Gram-Hadamard indequality we get:
\be |{\rm det} \tilde G^{h_v,T_v}({\bf t}_v)| \le
C^{\sum_{i=1}^{S_v}|P_{v_i}|-|P_v|-2(S_v-1)}\;.\lb{2.54a}\ee

By using \pref{fonfon},\pref{34},\pref{zza},\pref{zzaa} we get
\bea &&{1\over L\b}\sum_x \int dx_{0,v_0} |H_{\t,{\bf P},T,\a}(x,
x_{0,v_0})|\le c^{-n} [\prod_v {1\over S_v!}]
[ \prod_{v \in V_\chi} \g^{4 h_v S^{L}_v}
][\prod_{v\in H_\chi} \g^{(1+\b_v)(h_{\bar v'}-h_v) }][
\prod_{v\in V_\chi}\g^{-\a |P_v|  }]\nn\\
&&[\prod_{v\in V_\chi} \g^{-h_v(S^H_v+S^L_v-1)}]
(\sup_{k\ge h} |\n_k||)^{n_\n}|U|^{n_U}|\e|^{n_\e}\
\eea
where $\b_v=1$ if $v$ is a resonant cluster with more than 2 external lines.
%
%
Note that
\be
[\prod_{v\in V_\chi} \g^{-h_v(S^H_v+S^L_v-1)}]
[\prod_{v\in H_\chi}\g^{h_{\bar v'}-h_v}]\le \g^{h_{v_0}}
[\prod_{v\in V_\chi} \g^{-h_v (S^H_v+S^L_v)}]
[\prod_{v\in H_\chi}\g^{h_{\bar v'}}]\label{h1}
\ee
as $v_0\not
\in H_\chi$ so that $\prod_{v\in V_\chi} \g^{h_v}\le \g^{h_{v_0}}\prod_{v\not =v_0, v\in H_\chi} \g^{h_v}$.
Moreover
%
%
%
\be
[\prod_{v\in V_\chi} \g^{-h_v S^H_v}]
[\prod_{v\in H_\chi}\g^{h_{\bar v'}}]= 1
\ee
so that 
\be
[\prod_{v\in V_\chi} \g^{-h_v(S^H_v+S^L_v-1)}]
[\prod_{v\in H_\chi}\g^{h_{\bar v'}-h_v}]\le  \g^{h_{v_0}}
[\prod_{v\in V_\chi} \g^{-h_v S^L_v}]
\label{h1a}
\ee
We get
%
%
%
 \bea
&&{1\over L\b}
 \sum_x \int dx_{v_0}| H_{\t,\PP, T,\a}(x,\xx_{v_0})|\le\nn\\
 && \g^{h_{v_0}} [\prod_v {1\over S_v!}][ \prod_{v \in V_\chi} \g^{3 h_v S^{L}_v}]
[\prod_{v\in H_\chi} \g^{\b_v (h_{\bar v'}-h_v })]
[\prod_{v\in V_\chi}\g^{-\a |P_v|  }](\sup_{k\ge h} |\n_k||)^{n_\n}|U|^{n_U}|\e|^{n_\e}\
\eea
Note that $\sum_{\bf  P}[\prod_{v\in V_\chi}\g^{-\a |P_v|}]\le C^n$,
see for instance \S 3.7 of \cite{MM} for a proof; moreover 
$\sum_{\bf T}[\prod_v {1\over S_v!}]\le C^n$, see Lemma 2.4 of \cite{MM}.
The sum over the trees $\t$ is done performing the sum of unlabeled
trees and the sum over scales. The unlabeled trees can be bounded by $4^n$ by Caley formula, 
and the sum over the scales reduces to the sum over $h_v$, with $v\in V_\chi$, as
given a tree with such scales assigned, the others are of course determined. It remains to prove that
\be
\sum_{\{ h_v\}}][ \prod_{v \in V_\chi} \g^{3 h_v S^{L}_v}] [\prod_{v\in H_\chi} \g^{\b_v (h_{\bar v'}-h_v })]\le C^n\ee
We can write $\sum_{\{ h_v\}}=\sum_{h_v, v\in V_\chi \atop S^L_v\ge 1}+\sum_{h_v, v\in V_\chi\atop S^L_v=0}$; for the first sum we can simply use 
$[\prod_{v\in H_\chi} \g^{\b_v (h_{\bar v'}-h_v })]<1$ so that
\be\sum_{h_v, v\in V_\chi\atop S^L_v\ge 1}\prod_{v \in V_\chi} \g^{3 h_v S^{L}_v}\le C^n\ee
Regarding the second sum, we have to sum scales of vertices followed by
vertices $v_1,..,v_{S_v}$ which are all resonant.
We can still distinguish two cases; or $|P_{v_i}|=2$, $i=1,..,S_v$, that is the inner clusters of the cluster $v$ 
have two external lines, or not. 
\insertplot{890}{220}
{\ins{180pt}{150pt}{$h_v=h_{\hat v'}$}
\ins{190pt}{100pt}{$h_{\hat v}$}}%
{verticiT666}
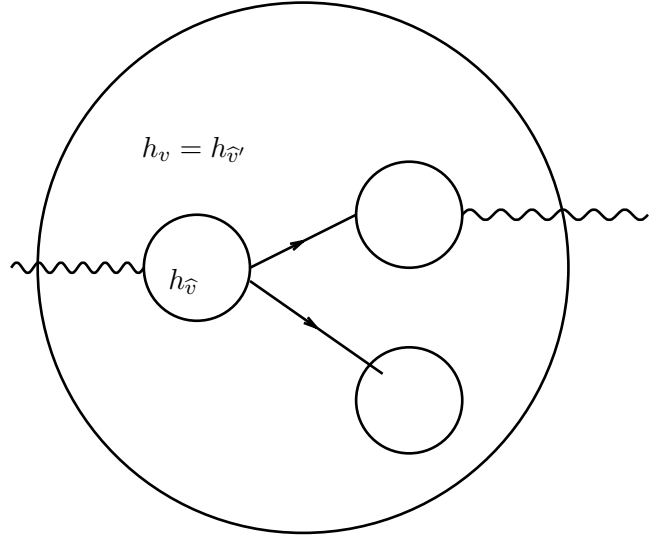
{\label{n9} A cluster $v$ with $S_v=3$, $S^L_v=0$; one inner cluster has more that 2 external lines.
}{0}
In the last case, there is surely a $j$ such that $|P_{v_j}|\ge 4$;
we call the scale of such inner cluster $h_{\hat v}$ and $\hat v'=v$, we can extract from the product $[\prod_{v\in H_\chi} \g^{\b_v (h_{\bar v'}-h_v }]$ a factor
$\g^{(h_{v}-h_{\hat v})}$ and we can use such factor to sum over 
$h_v\le h_{\hat v}$, see Fig.10. 
\insertplot{890}{220}
{\ins{180pt}{150pt}{$h_v=h_{\hat v'}$}
\ins{190pt}{100pt}{$h_{\hat v}$}}%
{verticiT777}
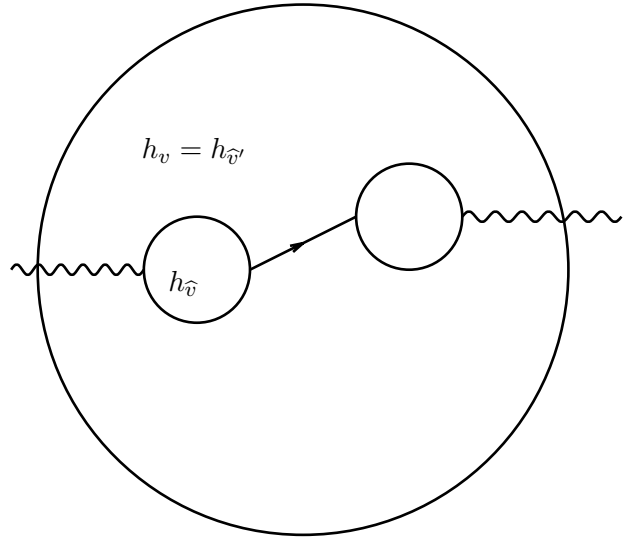
{\label{n9} A cluster $v$ with $S_v=2$, $S^L_v=0$; all inner clusters have two external lines}{0}
Otherwise, see Fig. 11,  the inner resonant clusters have all 2 external lines; therefore such clusters are connected by propagators with the same coordinate and momentum of the external lines, so that
there is no sum over $h_v$
as $h_{v}=h_{v'}+1$ 
by the support properties of the propagators.
\qed

Lemma 2.3 implies convergence of the expansion for the kernels of the effective potential, provided that
$\e, U$ and $\n_k$ are small enough; this last condition is ensured by choosing properly the counterterm $\n$
as a function of $\e,U$, as we will show below.

\subsection{Choice of the counterterm $\n$}

The above lemma ensures convergence provided that $\n_k$ are small for any $k$. We can write \pref{sper} as
\be
\n_{h-1}=\g\n_h+\sum_{n=2}^\io \b^{(h)}_n\label{ff}
\ee
\begin{lemma} If $n=n_\n+n_U+n_\e$ then $\b^{(h)}_n=0$ if $n_\e=n_U=0$; moreover
\be
|\b^{(h)}_n|\le C^n (\sup_{k\ge h} |\n_k||)^{n_\n}|U|^{n_U}|\e|^{n_\e}\label{bra2}
\ee
\end{lemma}

{\it Proof} By \pref{sper}
\be 
\g^h\b^{(h)}_n=\sum_{\t\in \TT_{h,n}} \sum_{T\in {\bf T}}
 \sum_{{\bf P}\in {\cal P}_\t}\sum_{\a\in A_T}
\int dx_{0,v_0} H_{\t,{\bf P},T}(0,
x_{0,v_0})\label{sper1}
\ee
and $v_0\in V_\chi$. If $n_\e=n_U=0$ the only contributions is from chain graphs, whose value is given by a product of $\hat g^{k}(x',k_0)$, $k\ge h$ computed at $k_0=x'=0$, hence they are vanishing by the compact support properties of the propagators  $\hat g^{k}(0;0)=0$.
Moreover the r.h.s. of \pref{sper1}
verifies the same bound as the r.h.s. of \pref{bra1} with $\g^h$ replacing $\g^{h_{v_0}}$ as
$h_{v_0}=h$ as $v_0\in V_\chi$ because $\LL\RR=0$. 
\qed
\vskip.5cm
It remains to prove that we can choose $\n$ so that $\n_h$ is small. First we write
\be
\n_{h}=\g^{-h+1}(\n+\sum_{k=h+1}^1 \g^k \b_k)
\ee
We introduce a sequence of $\n_k^{(n)}$ such that $\n_h^{(0)}=0$ and 
\be
\n_{h}^{(n)}=\g^{-h+1}(\n+\sum_{k=h+1}^1 \g^k \b_k^{(n-1)})
\ee
where $\b_k^{(n-1)}$ is obtained from $\b_k$ replacing $\n_k$ with $\n_k^{(n-1)}$.
\begin{lemma} Setting 
\be
\n^{(n)}=-\sum_{k=h+1}^1 \g^k \b_k^{(n-1)}
\ee
then the sequence $\n_h^{(n)}$, $h\le 1$, $\n_1\equiv \n$, converges uniformly to $\n_h$ with $|\n_h|\le C \max(|\e|,|U|)$
\end{lemma}

{\it Proof} We show by induction that
\be
|\n_h^{(n)}-\n_h^{(n-1)}|\le C (\max (|\e|,|U|))^n \quad\quad
|\b_h^{(n)}-\b_h^{(n-1)}|\le C (\max (|\e|,|U|))^n\label{ind}
\ee
For $n=1$ then $\n^{(1)}_h=0$ for $h\le 1$ and $\n^{(1)}_1$, 
by \pref{bra2}, is such that $|\n_1^{(1)}|\le C \max(|\e|,|U|)$, for $\e, U$ small enough. 
If $n > 1$ then $\b_h^{(n)}-\b_h^{(n-1)}$
can be written as sum of terms in which there is at least a 
$\n_h^{(n)}-\n_h^{(n-1)}$, hence \pref{ind} follows by \pref{bra2}, as $|\b_h^{(n)}-\b_h^{(n-1)}|\le \bar C
 \max(|\e|,|U|) |\n_h^{(n)}-\n_h^{(n-1)}|$ as $\b_n^{(h)}$ is vanishing if $n_\e=n_U=0$.
Therefore uniform convergence follows.
\qed

\subsection{The 2-point function}

We have finally to get a bound for the two-point function, which can be written as
\be
S(\xx,\yy)=\sum_{n=2}^\io H_n(\xx,\yy)
\ee
where $H_n(\xx,\yy)$ is sum over trees with $n$ end-points and any value of $h_{v_0}$, among which there are $2$ special end-points associated to the external lines and $n-2$ are associated normal end-points of type $\e,U,\n$. Note that 
there is necessarily a path $c_{w_1,w_2}$ in $\hat T_v$ connecting the points $w_1$, with $\xx_{w_1}=\xx$
and $w_2$ with $\xx_{w_2}=\yy$ such that by \pref{fa} $|x-y|\le  |c_{w_1,w_2}|$;
moreover $|c_{w_1,w_2}|\le n$ so that $H_n=0$ for 
$n< |x-y|$. 
Therefore with respect to the bound to the effective potential 
\pref{bra1} there is an extra $\g^{h_{v_0}}$ for a missing integral due to the fact that $\xx,\yy$ are fixed and
and extra $\g^{-2h_{v_0}}$ for the presence of the external lines. The sum over the scales is bounded by $|\bar h|$ with
\bea
&&\g^{-\bar h}\le \max_{k\in 0,n}\max_{\r=\pm 1}{1\over ||\o (x+k)-\o\r\hat x-2\d_{\r,-1}\th||}\le\nn\\
 &&C (1+\min\{|x|,|y|\}+n)^\t\le C (1+\min\{|x|,|y|\})^\t(1+{n\over 1+\min\{|x|,|y|\}})^\t
\eea
so that in conclusion
\bea
&&|S(\xx,\yy)|\le \sum_{n\ge |x-y|}(\max(|\e|,|U|))^n C^n \log [(1+\min\{|x|,|y|\})^\t(1+{n\over 1+\min\{|x|,|y|\}})^\t]\nn\\
&&\le e^{-{\a\over 2}|\log max(|\e|,|U|)||x-y|}
\log [(1+\min\{|x|,|y|\})^\t]
\eea
We can get another bound, which is better for large $|x_0-y_0|$;
by integrating by parts and using that each derivative carry an extra $\g^{-h_{v_0}}$ one gets 
\be
|S(\xx,\yy)|\le e^{-{\a\over 2}|\log max(|\e|,|U|)||x-y|}{C_N\over 1+(\min\{|x|,|y|\}^{-\t}|x_0-y_0|)^N}
\ee
and combining the above two bounds, Theorem 1.1 follows.

\end{document}